\documentclass[twocolumn,english,aps,superscriptaddress,pra]{revtex4-1}

\usepackage{xcolor}
\usepackage{array}
\usepackage[latin9]{inputenc}
\usepackage{amsmath}
\usepackage{amssymb}
\usepackage{graphicx}
\usepackage{babel}
\usepackage{mathrsfs}
\usepackage{amsfonts}
\usepackage{epstopdf}
\usepackage{float}
\usepackage{color}
\usepackage{natbib}

\begin{document}
\title{Emergent Symmetries and Interactions in High Dimensions: \\ An Isolated Fixed Point Versus A Manifold of Strongly Interacting Fixed Points }

\author{Fei Zhou}

\affiliation{Department of Physics and Astronomy, University of British Columbia, 6224 Agricultural Road, Vancouver, BC, V6T 1Z1, Canada}

\begin{abstract}
In this article, we explore conditions of continuous emergent symmetries in gapless states, either as topological quantum critical points (TQCPs) or a stable phase with protecting symmetries and 
connections to smooth deformations of the gapped states around.
For a wide class of gapless states that can be associated with fully-isolated scale invariant fixed points, we illustrate that there shall be emergent continuous symmetries that are directly related to smooth deformations of gapped states with symmetries lower than the  protecting ones $G_p$.
A short distance invariance of gapped states under deformations can descend to be an emergent continuous symmetry when approaching the gapless limit. 
Around one TQCP with $G_p=Z^T_2$ symmetry, we construct these deformations explicitly and show emergence of symmetries via fully gapped quaternion superconducting states that break the protecting symmetry $G_p=Z^T_2$. 
For a 3D TQCP in DIII classes with $G_p=Z^T_2$, $U_{EM}=U(1)$ and $N_f=\frac{1}{2}$ fermions but without charge $U(1)$ symmetry, we further explicitly construct a corresponding boundary representation based on a $4D$ topological state with lattice symmetry $H=Z^T_2 \ltimes U(1)$ and $N_f={1}$ fermions. The lattice model is shown to be dual to a conventional $4D$ topological insulator.
Although emergent continuous symmetries appear to be robust at weakly interacting TQCPs, we further show the breakdown of such one-to-one correspondence between deformations of gapped states and emergent continuous symmetries when gapless states become strongly interacting. In a strongly interacting limit,
gapless states can be represented by a smooth manifold of conformal-field-theory fixed points rather than a fully isolated one. A smooth manifold of strong coupling fixed points hinders emergence of a continuous emergent symmetry in the strongly interacting gapless limit, as deformations no longer leave a gapless state or a TQCP invariant, unlike in the more conventional weakly interacting case. This typically reduces continuous emergent symmetries to a discrete symmetry originating from duality transformations under the protection symmetry $G_p$.

\end{abstract}

\date{\today}

\maketitle

\section{Introduction}

Recently, there have been quite fascinating discussions on gapless topological states\cite{Ji20,Ji22,Verresen21,Thorngren21,Scaffidi17,Verresen18,Chatterjee22,Wen17,Baum15,Fidkowski11a,Sau11,Ruhman15,Jiang18,Keselman15,Iemini15,Kainaris15}. Unlike gapped topological states, a subject which remarkable progress had been made on during the last decade or so, and become reasonably well understood\cite{Schnyder08,Kitaev09,Qi11,Bernevig13,Hasan10,Qi10,Fidkowski10,Fidkowski11,Chen10,Chen12,Chen13,Hastings05,Fidkowski13,Metlitski15,Wang15,Song17}, gapless states remain to be a subject of intensive research, especially when it comes to their general structures. In our view, there are a few major challenges when topological gapless states are concerned. One has been that the gapless low energy sector is much richer in terms of excitations compared to fully gapped states.
 So for instance the topological entanglement entropy and topological degeneracy that are commonly used as nonlocal topological order of gapped phases can not be clearly defined\cite{Levin06,Kitaev06}, at least not possible without major modifications. The other challenge which is a closely related one but more from practical observation points of view is that surface states as hallmark signatures of most topological states, especially symmetry protected ones, typically become faded away when gaps shrink to zero and surface states in many cases can no longer be well isolated from the gapless bulks. This standard paradigm however can be violated in some critical states. 
Recent studies concentrated on one dimensions have identified a few examples of surprising edge modes in gapless conformal-field-theory states\cite{Jiang18,Scaffidi17,Verresen21,Thorngren21}.

Perhaps the one very surprising aspect of gapless states, say as quantum critical ones, is that their symmetries can be much higher than gapped topological phases in the close proximity with protection symmetry $G_p$.  Here we use the subscript in $G_p$ to refer to the protecting symmetry of topological states.
Topological quantum critical points (TQCPs) separating topologically different phases with protection symmetry $G_p$ can further process emergent symmetries which had been widely observed. Existence of these TQCPs solely rely on 
surrounding gapped topological phases protected by symmetry $G_p$ and are well-defined only when symmetry $G_p$ is present. In fact, the presence of protecting symmetry $G_p$ is a necessary condition for these TQCPs although it is not sufficient.

Furthermore, generally we believe these emergent symmetries lead to t'Hooft anomalies when we further impose them locally or gauge them. In very broad classes of studies, one can show that the t'Hooft anomalies in $d$-dimension gapless systems
with emergent symmetries can be related to difficulties of gauging $d$-dimension surfaces or boundaries of a $(d+1)$-dimension topological bulk. This feature of topological matter had been utilized to classify quantum anomalies in quantum gauge field theories\cite{t'Hooft76,Adler69,Bell69,Wen13}.
All these appear naturally in the context of effective field theories (EFT), although at the moment it is
not entirely clear what happens at a more microscopic level or at intermediate or ultraviolet scales where simple {EFT}s may not be adequate.  

If gapless states are in a stable phase protected by a symmetry group $G_p$, they shall be represented by infrared stable fixed points with the protection symmetry $G_p$. Even such gapless states usually have higher emergent symmetries in addition to protecting symmetry $G_p$. As far as they are not gapped by breaking the protecting symmetry $G_p$, there appears a robust infrared emergent symmetry group, even robust again weak interactions, at least in the infrared limit. 

While many observations had been made on emergent symmetries and generalized emergent symmetries have even applied to classify gapless phases, 
a few important questions appear to have not been answered so far. 

{\bf Q.a}. One is to what extend one can adopt the concept or the paradigm of emergent symmetries for discussions of physical phenomena or measureables? Can one actually directly probe such emergence in measurements? Answer to this question remains largely unknown although there are limited evidences pointing a positive answer. 

{\bf Q.b}. The other general question is the origin of these puzzling emergent symmetries, why they usually show up at TQCPs in topological states with protecting symmetries but not in most conventional order-disorder transitions or where they are from, how they are related to gapped state properties and how they look like at a more  microscopic level or in lattice models. We are not aware of intensive discussions on this general aspect but believe it is important enough to deserve more efforts to have clear understanding and offer transparent and intuitive answers. Specifically, we are interested in understanding emergent continuous symmetries in high dimensions and whether and how they can be related to deformations of gapped phases
around TQCPs.

{\bf Q.c}. The third important question is perhaps how robust emergent symmetries in gapless states are and to what extent they depend on weak or strong interactions and types of fixed points that are involved at TQCPs.

This article is devoted to partially answering the second and third questions above with emphases on smooth deformations of ground states, how deformations look like in the surface-state representation of a $(d+1)$-dimension lattice model,
and the roles of strong interactions. The phenomenological picture emerging from our studies bridges a gap between microscopic wavefunctions, lattice models and EFTs, and between weakly interacting limits and strongly interacting ones.
Although most quantitative analyses will be carried out in the context of DIII class states of topological superfluids that we have been working on quite intensively during the last few years, the main phenomenology and approach appear to be general enough to be applicable to other symmetry protected topological states. 

The main conclusion is that it is $G_p$, the protecting symmetry itself, which is always required for defining topologically states, makes gapless TQCPs more symmetric than they ought to be. Same shall even be applicable to symmetry protected gapless {\em phases}. By contrast, generic order-disordered phases are much more robust against variations of $G$, the symmetry group of interaction Hamiltonians and condensation itself can often occur disregarding symmetry groups although the Wilson-Fisher universality does crucially depend on symmetry groups $G$.

The article is organized as follows.
In Section II, we make a few general remarks and clarifications of our approach and issues studied in this manuscript.
In Section III, we discuss possibilities of deforming gapped symmetry protected topological (SPT) states by lowering or breaking the protecting symmetry group $G_p$ so to smoothly connect topologically distinct states.
We  demonstrate that such general deformations, which shall exist, can lead to an invariance when performing on a surrounded gapless state and therefore result in continuous emergent symmetries. 
The sufficient condition for this to happen is that the TQCP is associated with an isolated scale invariant fixed point and is disconnected with other fixed point states.
Here, we also contrast explicitly to what happens in an order-disorder quantum critical point
where such deformations are generically forbidden because of robustness of condensation of bosons. 

This part of discussions is presented in terms of general topological quantum critical point phenomenology and we strongly believe shall be applicable to many other symmetry protected gapless states beyond what we will focus later in this article.

In Section IV, we summarize our main results on the relation between smooth deformations and emergent continuous symmetries when a TQCP with protecting symmetries is either represented by an isolated fixed point or lives in a manifold of fixed points.
We illsutrate the one-to-one correspondence in the former case while in the later case such a relation generally breaks down. 
In Section V, we show explicit deformations in a quaternion superconductor and illustrate such deformations of wave-functions can be applied to smoothly connect two topologically distinct DIII class time-reversal-symmetric topological superconducting states.  
We also apply the similar idea to a gapless nodal point phase protected by a reflection parity symmetry. In both cases, we show the explicit connection between deformations of gapped states with no protecting symmetries and higher emergent symmetries in gapless states $H=G_p\ltimes U_{EM}$, where $U_{EM}$ is the emergent symmetry induced by a deformation group $U_D$. In Section VI, we further treat solid state ground states as Vacua of effective field theories and illustrate such deformations beyond the wave-function approach to include generic interactions between emergent real fermion fields. 

We further explicitly construct $(d+1)$-dimension lattice models with higher protecting symmetries including the emergent symmetry of the gapless states, $H=G_p \ltimes U_{EM}$. 
We demonstrate that the surface states of $(d+1)$-dimensional topological states protected by the larger symmetry $H$ precisely project out $d$-dimension gapless states or a TQCP.
The opposite surfaces represent two TQCPs with the same protection symmetry $G_p$ and emergent symmetry $U_{EM}=U_D$ and they are connected by a parity transformation.
For $d=3$ with $U_{EM}=U(1)$ and a TQCP in DIII class superconductors without charge-$U(1)$ symmetry, such a $(d+1)$-dimension lattice model is dual to a $4D$ topological insulator with charge $U(1)$ symmetry and its surface states form a representation of TQCPs separating two gapped topologically distinct states with protecting symmetry $G_p$.
Details are presented in Appendix D.

In Section VII, we discuss a strongly coupling limit where a TQCP is identified to be a fixed point in a smooth manifold of other fixed points without protecting symmetry $G_p$. We show in this limit the deformations can trace out other conformal fixed points without protecting symmetry but don't leave the TQCP invariant. There are no continuous emergent symmetries unlike in the weakly interacting limit where the fixed point is an isolated one. Instead, the gapless TQCP can only have a discrete $Z_2$ parity symmetry. In Section VIII, we conclude and discuss open questions.

Before leaving the introduction, we also want to point out that emergent gauge symmetries and/or gauge fields had been previously emphasized in other beyond-Landau paradigm quantum critical points.
For instance, symmetries can emerge at deconfined quantum critical points\cite{Senthil04} that separate an N$\acute{e}$el state 
which breaks $SU(2)$ spin rotation and translational symmetries, and a valence bond solid phase which breaks lattice translational and rotational symmetries but with spin rotation symmetry unbroken. This article on the other hand will be mainly concerned with emergent symmetries in gapless states separating different topological states that are invariant under the same protecting symmetry group $G_p$, without lattice translational symmetry breaking.

 \section{Three general remarks on our approach to emergent symmetries and beyond}

Before starting our discussions on the subject, we would like to make three general remarks on the method employed in this article, the relations between our conclusions and a few previous results, and the limitation of our current studies.

{\em Remark One}:

To understand  {\bf Q.b} raised above, the origin of these puzzling emergent symmetries that usually show up at TQCPs with protecting symmetries but not in the more conventional Landau order-disorder transitions, 
we find it is extremely productive to analyze {\em the phase diagram (or boundary) topology } in the Landau paradigm and compare it with that around the TQCPs. 
In the Landau paradigm, phases with different off-diagonal long range order can not be smoothly deformed into each other even when the symmetry group of the Hamiltonian $G$ is lowered into a smaller group $G_s$.
One of the productive way to proceed is to examine whether this is still true when SPT phases separated by TQCPs are concerned. 
For this purpose of probing the phase boundary topology around a TQCP, we study the deformation of gapped phases around TQCPs to see if they can be smoothly connected when the protecting symmetry
$G_p$ is explicitly broken. {\em Here it is imperative that the deformation group breaks the protecting symmetry $G_p$ as otherwise it can never smoothly connect two topologically distinct SPT states.}
It turns out that the distinct feature of deformable gapped states around a TQCP plays a very important role in the emergent symmetry---the emergent symmetry is defined by the same group as the deformation one.

The deformation group which allows us to smoothly connect different SPT phases also plays an instrumental role in our discussions on {\bf Q.c}, whether the emergent symmetry group depends on interactions and how.

It is worth emphasizing that the deformation group itself has to break the protecting symmetry $G_p$ in order to connect two gapped phases separated by a TQCP, according the general paradigm of SPT\cite{Chen10}. 
The path we are following here towards the emergent symmetry in gapless states is partially motivated by the deformation construction in gapped phases in Ref.\cite{Hastings05,Chen10} but differs in detailed structures.
And here we only consider the deformation group generated by a local operator which we then naturally relate to a standard symmetry group.

{\em Remark Two}:

In a previous study of SPT phase transitions, emergent gauge fields such as Non-abelian deconfined fields related to Banks-Zaks fixed points have been proposed as effective field theories of an exotic class of TQCPs\cite{Bi19}. 
There, the authors took a top-to-bottom approach asking what phase transitions in SPTs can be described by a class of no-abelian gauge fields interacting with fermions if TQCPs are deconfined critical points.

The current article is mainly concerned with the issue of emergent symmetries. And we instead take a {\em bottom-to-up} approach asking what shall be the emergent symmetries at TQCPs in SPTs with a given protecting symmetry and a given change of topologies. In  the class of TQCPs in DIII class topological superconductors we have investigated, because of the t'Hooft anomalies, we also illustrate that there can be no emergent gauge fields at TQCPs.
So these TQCPS studied in this article do not belong to the universality classes proposed in Ref.\cite{Bi19}.

In this article, we are mainly focused on a fundamental representation of TQCPs within SPT states protected by symmetry $G_p$, unique and generic for quantum phase transitions with a {\em minimum change} in topological invariants that itself is set by the symmetry $G_p$. 
In general, we find the emergent symmetry as a group shall be a function of both protecting symmetries and changes of topological invariants and so there can also be emergent gauge fields. The general conditions of a) when there shall be emergent 
gauge fields and b) what kinds of gauge fields shall emerge, given i) the protecting $G_p$ and ii) a specific change of global topologies are a fascinating question we don't have an answer yet and perhaps shall be more thoroughly studied in the future.

{\em Remark Three}:

Some transitions in SPT in low dimensions can also be mapped into transitions involving symmetry breaking via a non-local transformation such as Jordan-Wigner transformation \cite{Fidkowski10}. 
Recent studies in connections to quantum information even suggest that one can obtain a topologically ordered (TO) state from a SPT by following a measurement protocol which effectively performs a non-local 
Kramer-Wannier transformation\cite{Nat21}. 

If one further includes non-local transformation, it becomes more challenging to clearly define phase boundaries between different quantum matter and/or TQCPs, than in a more standard frame work where we usually restrict ourselves to local unitary transformations or local deformations.  We do not have definitive answers to fermionic TQCPs in the presence of non-local deformations, especially in high dimensions, but speculate that to discuss emergent symmetries near TQCPs at that general level, one needs to carefully take into account higher form symmetries of fermions (also see discussions in conclusion) which is beyond the scope of the current article. But this is definitely a very fascinating issue to pursue in the future: How to identify high form emergent symmetries near fermonic TQCPs?

However, if we just take a specific one-dimensional example in Ref. \cite{Fidkowski10} where a mapping between the single quantum Ising chain and a BDI class superconducting chain had been established, then we can explicitly show that the TQCP 
in the BDI class in the single chain limit doesn't exhibit a {\em continuous} emergent symmetry. That is consistent with the usual identification of a $Z_2$ emergent 0-form symmetry at a quantum critical point in a 1D Ising spin chain, due to duality. 
As our studies mainly explore a relation between continuous deformation groups of gapped states and continuous emergent symmetries appearing at TQCPs, our approach in this case only states that if there are emergent symmetries,
they have to be discrete ones which is fully consistent with the single chain Ising model and related BDI class physics.

Finally, let me emphasize that all the discussions of emergent symmetries in this article are related to fixed points and the infrared physics indicated by these fixed points. 
The emergent symmetries can be broken when moving away from the fixed points but only in a quite subtle way. In general, irrelevant operators do lower the emergent symmetry if the energy scale is increased but the emergent symmetry shall remain in the IR limit as an asymptotic symmetry. What is demonstrated in this manuscript is that there are TQCPs with strong interactions where the emergent symmetry in the IR limit is distinctly different from that at an isolated fixed point.

\section{Emergent continuous symmetries and surrounding gapped states without Protecting symmetries}

In this section, we will connect infrared emergent symmetries in a gapless state to an ultraviolet invariance of the ground state wave-functions of surrounding gapped states without protecting symmetries when a symmetry transformation is carried out in these states. Our discussions on this connection will be focused on two different but related general aspects of the phenomena.

\noindent  A) Paths of continuous deformation of symmetry protecting topological states into surrounding gapped states without protecting symmetries, can naturally lead to continuous emergent symmetries in gapless states as generically these paths or corresponding transformations can be contractable when acting on a gapless state. Existence of such paths, without encountering a line of critical points, on the other hand is always suggested or generally granted by the symmetry protecting nature of topological states.

\noindent B) Conversely, the emergent symmetries and corresponding transformations in the gapless state always imply an ultraviolet invariance of surrounding gapped states when performing the same transformations. Furthermore, the local operators in these gapped states that are non-scale invariant nevertheless shall all have the same scaling dimensions at short distances and are in the same universality class as the gapless state. And they naturally belong to the same phase from the scaling points of view.
Paths of gapped states traced out by the transformations can not cross a line of critical points which otherwise would lead to different scaling behaviors.

\subsection{Topological states with protecting symmetries and adiabatic deformation without protecting symmetries}

\begin{figure}
\includegraphics[width=6cm]{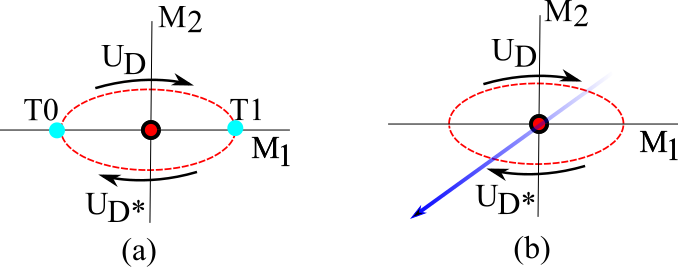}
\centering
\caption{ \footnotesize a) Deformations of gapped states around a gapless TQCP with protection symmetry $G_p$. Only the horizontal axis $M_1$ is protected by $G_p$ but the rest of the plane where $M_2$ breaks the symmetry $G_p$ has a lower symmetry.
Smooth deformations $U_D(\lambda)$ along the upper half of the dashed oval connects two topologically distinct gapped states $T0$,$T1$(light blue dots) located along the horizontal axis $M_1$. The lower half of the dashed oval describes a deformation $U_{D*}(\lambda)$ induced when applying $G_p$ to $U_D(\lambda)$, or $U_{D*}=U^{-1}_{G_p}(\lambda) U_D U_{G_p}(\lambda)$. 
$U_D$ together with $U_{D*}$ trace out a closed loop (dashed oval) around the TQCP leading to an emergent symmetry at the TQCP.  b) Loop (dashed oval) traced out by smooth deformations of gapped states around a stable gapless state (red dot in the centre) protected by a symmetry $G_p$.
Symmetry $G_p$ is only present along the third axis
perpendicular to $M_1$-$M_2$ plane (the arrowed axis from the red dot in the centre) where gapless states form a stable phase. The 3D volume away from the arrowed axis breaks the symmetry of $G_p$ and states are fully gapped except along the centre arrowed axis.}
\label{TQCPEMS}
\end{figure}

We start with the aspect A and focus on gapped states next to a gapless state of interest. Although the discussions can be easily generalized to a gapless phase, for simplicity of our discussions, we assume the gapless state is quantum critical and represents a topological quantum critical point (TQCP). A TQCP separates two topologically distinct phases with protecting symmetries $G_p$. We can introduce an axis of mass operators with protecting symmetries $G_p$ and TQCP is an isolated point along the axis of $G_p$ with the mass set to be zero (See Fig.\ref{TQCPEMS}).
 
The gapped topological ground state is invariant under $H=G_p$ and so has a non-degenerate manifold $\mathcal{M}_g=G_p/H=1$. It is critical that in our discussions, $\mathcal{M}_g$ is the same on two sides of a TQCP; i.e. two phases only differs by topologies not in terms of the standard symmetries $H$.

As a side note, let us mention that $G_p$ in general can be different from $G$, the symmetry group of the underlying interacting Hamiltonian. For instance for a topological insulator of time reversal invariance, $G_p=Z^T_2 \ltimes U(1)$; in this case,
$G_p=G$ without conventional symmetry breaking and $\mathcal{M}_g=1$. However when applying to time reversal topological superfluids in DIII class states, $G_p=Z^T_2$ without $U(1)$ symmetry but $G=Z^T_2 \ltimes U(1)$. The spontaneous symmetry breaking leads to 
$\mathcal{M}_g=G/H$ where $H=G_p=Z^T_2$ is the invariant group of the ground state. In this convention, $\mathcal{M}_g=U(1)$. In our later discussions, we follow the notion used in the community of topological matter by working with the protecting symmetry $G_p$ directly, and $\mathcal{M}_g=G_p/H$ rather than with $G$ and the difference between these two presentations does not play a role in our discussions on topology. 

We start with a gapped topological state $|\Psi_{T0} \rangle$ with protecting symmetries $G_p$. The topological phase is well defined only when $G_p$ symmetry is present in the system. However, if we allow deformations via {\em local unitary transformations} into states without $G_p$, then symmetry protection is absent and the initial state $|\Psi_{T0} \rangle$ can always be smoothly connected to trivial states or other topologically distinct states on the other side of TQCPs via these deformations.
Let us name the final state $|\Psi_{T1} \rangle$. The unitary transformation that connects them can be specified as

\begin{eqnarray}
|\Psi_{T\lambda} \rangle &=& U_D(\lambda \in [0,1]) | \Psi_{T0} \rangle  \nonumber \\
|\Psi_{T0} \rangle &=&| \Psi_{T\lambda=0} \rangle, |\Psi_{T1} \rangle=  |\Psi_{T\lambda=1} \rangle
\label{deform}
\end{eqnarray} 
The deformation or transformation $U_D(\lambda)$ traces out a path of states, $\mathcal{C}, $ that connects the initial $|\Psi_{T0} \rangle$ to the final state $|\Psi_{T1} \rangle$ when $\lambda$ varies continuously from zero to one.

For our purposes of discussions later, we restrict ourself to unitary transformation induced by by an operator of the form $\sum_{\bf r}  \mathcal{O}(\bf r)$ where $\mathcal{O}({\bf r})$ is local, the correlation functions of local operators shall be invariant under such deformation. That is if

\begin{eqnarray}
[\mathcal{O} ({\bf r}), \mathcal{O}({\bf r}')]=0 & \mbox{ if ${\bf r}\neq {\bf r}'$.} 
\end{eqnarray}
the states traced out by $U$ shall be correlated in the same way.

This indicates two important features of the deformation. First, all the states including $|\Psi_{T1}\rangle$ are short ranged correlated or fully gapped just as the initial state $|\Psi_{T0}\rangle$. 
The path traced out by $U(\lambda \in [0,1])$ can not cross a line of critical points where states are fully scale and conformal symmetric.

Second, some of the states along path $\mathcal{C}$ (excluding the initial and final), if not all the states, must not have the protecting symmetry $G_p$. Otherwise, a path induced by $U_D$, fully protected by symmetry $G_p$, 
shall remain entirely in a given gapped topological phase and can't connect states in two topologically distinct phases. And $\mathcal{C}$ can't cross a TQCP as local unitary transformations preserve short range correlations.  

As below we are  mainly interested in the case of DIII class states with time reversal symmetry, we will focus the case when $G_p=Z^T_2$.
Since states along $\mathcal{C}$ are not invariant under protecting symmetry group $G_p$ and all are fully gapped, let us assume that $\mathcal{C}$ transforms into $\mathcal{C}_{*}$ under the action of $G_p$ or $U_{G_p}$.
The corresponding anti-unitary transformation to produce such a path is

\begin{eqnarray}
U_{D*}(\lambda) =U_{G_p} U_D(\lambda ) U^{-1}_{G_p}, \lambda  \in [0,1].
\end{eqnarray}

We further notice that the gapped topological states $|\Psi_{T0,T1}\rangle$ shall be invariant under the group of protecting symmetry  $U_{G_p}$,
\begin{eqnarray}
U_{G_p} |\Psi_{T0,T1} \rangle=|\Psi_{T0,T1} \rangle.
\label{symmetry}
\end{eqnarray} 

Eq.\ref{symmetry} together with Eq.\ref{deform} indicate that at $\lambda=0,1$, 
\begin{eqnarray}
&& U_{G_p} U_D(\lambda=1) U^{-1}_{G_p} = U_D(\lambda=1); \\
&& U_{G_p} U_D(\lambda=0) U^{-1}_{G_p} =U_D(\lambda=0)
\label{UTRS10}
\end{eqnarray}
while at $\lambda \neq 0,1$, the above equalities do not hold.

Putting all the considerations together, we then conclude that there shall be continuous deformation given by a closed path $\mathcal{C} +\mathcal{C}_*$ that precisely encloses the gapless state in the Hilbert space. The corresponding transformation is defined as $\tilde{U}(\lambda \in [0,2])$,

\begin{eqnarray}
\tilde{U}_D(\lambda \in [0,2]) = \left\{ \begin{array}{cc} U_D (\lambda) & \mbox {if $\lambda \in [0,1]$}; \\ 
U_{D*}(2-\lambda) & \mbox{if $\lambda \in [1,2]$.}
\end{array}\right.
\end{eqnarray}
where $U_D(0)=U_{D*}(0)$ and $U_D(\lambda=1)=U_{D*}(\lambda=1)$ following Eq.\ref{UTRS10}.

To summarize, we find that since two gapped states belonging to two different topological phases can be deformed into each other without encountering a line of critical points when the protecting symmetry $G_p$ is explicitly broken, there shall be a closed path $\mathcal{C}_T=\mathcal{C} +\mathcal{C}_*$ connecting the two states. The path traces out gapped states without protecting symmetries but with two point ($\lambda=0, 1$) along the path held fixed in two topological phases with protecting symmetry $G_p$(See Fig.\ref{TQCPEMS}).

The size of the path $\mathcal{C}_T$ can be adjusted by choosing $|\Psi_{T0} \rangle$ or $|\Psi_{T1} \rangle$. This is effectively to set the mass gap under deformations. The key point is that as the path encloses a TQCP, it is not contractable unless
it shrinks into the TQCP, the gapless limit. And as the local unitary transformation leads to the same scaling property of correlations, local operators in all the states along the path shall have the same universal scaling dimension.
This also implies that all gapped states remain invariant in an ultraviolet limit under the transformation of $U_D(\lambda)$.

This shall be the same universal scaling dimension appearing at the TQCP.  By choosing $|\Psi_{T0, T1} \rangle$ to be very close to a TQCP, the close path has to shrink toward the TQCP. All the universal structures in gapped states along the path eventually descend down to the infrared limit when states turn into
gapless and critical. Specifically, the unitary transformations along the path $\mathcal{C}_T$ can now directly act on a TQCP if the loop of $\mathcal{C}_T$ shrinks down to zero size.

If the TQCP is the only gapless state with long range correlations surrounded by fully gapped states (without the protecting symmetries), the transformations 
have to leave the TQCP invariant as the local transformation can not turn it into gapped states. Hence under such an assumption, we have established a firm relation between the closed deformation paths tracing out gapped states in the absence of the protecting symmetries $G_p$ and emergent symmetries at a gapless
TQCP. This simple observation turns out to point to an origin of emergent symmetries that we have frequently encountered in many gapless states related to symmetric protected topological phases.

Before leaving this subsection, we emphasize that the the equivalence between the above smooth deformations of gapped states without protection symmetries $G_p$, and emergent continuous symmetries in a TQCP separating symmetry protected topologically distinct phases, relies crucially on the isolation of the fixed point, i.e. the TQCP is represented by {\em a completely isolated fixed point}. 
An example of such is a free fermion fixed point which is applicable to all TQCPs in symmetry protected topological phases in the ten-fold-way classification\cite{Schnyder08,Kitaev09}.
In high dimensions, the free fermion fixed point is further infrared stable in terms of interactions.

However, the one-to-one correspondence between smooth continuous deformations of gapped states and emergent symmetries in the gapless TQCP would break down if there were a smooth manifold of fixed points to which a TQCP with protecting symmetry belongs to. In that case, the unitary transformations could simply lead to deformations of a TQCP, or a fixed point with symmetry $G_p$, into other fixed points in the manifold typically with lower symmetries rather than leaving the TQCP invariant. 
Later on in Section VII , we will illustrate that this indeed happens if the gapless TQCP is {\it not isolated} and there are a family of infrared stable strongly coupling conformal fixed points that can be traced out by continuous deformations.
 
Another way to put this sufficient condition for emergent continuous symmetries is that {\em the gapless state at TQCP has to be a singlet under the action of deformations $U_D$}. And if the scale invariant fixed point for the TQCP is a fully isolated one even when the protecting symmetry is lowered, then this is sufficient to keep the gapless state being symmetric under the deformations. However, if the fixed point belongs to a larger smooth manifold of fixed points, then the deformations will trace out a path of scale invariant fixed points in the smooth manifold with lower symmetries.

\subsection{Implications of emergent IR symmetries on UV invariance in gapped states without protecting symmetries}

In the previous subsection, we have illustrated how smooth deformations among gapped states without protecting symmetries that generally have to exist  can be applied to connect two topologically distinct symmetry protected states.
And the local transformations or smooth deformations eventually result in emergent symmetries at a TQCP or in gapless states.
In this subsection, we take a look at the issue with an opposite view and ask what can or have to happen when the unitary transformations $U_D(\lambda)$, $\lambda \in [0,2]$ that leave the gapless states invariant and define the emergent  continuous symmetry groups $U_{EM}$ are applied to a gapped topological state $|\Psi_{T0} \rangle$. The scenario is fully consistent with the discussions presented in the previous section. That is a symmetry protected topological state has to be smoothly connected to a topologically distinct state in another phase under the protection of same symmetry group $G_p$ without crossing a line of critical point. The gapped states traced out under $U_D(\lambda)$ or by the deformation path $\mathcal{C}_T$ have to break the protection symmetry $G_p$.

First all, a generic element in group $U_D(\lambda)$, doesn't leave $|\Psi_{T0} \rangle$ invariant as $U_D$ represents an emergent symmetry in gapless states rather than a lattice symmetry applicable to all gapped states with protecting symmetry $G_p$. So it has to trace out a path along which all gapped states have the same correlations. But if such a path defined by $U_D$ were entirely within a single gapped topological phase with protection symmetry $G_p$ centred around a fully gapped state, then when the $|\Psi_{T0} \rangle$ approaches the center,
the path shrinks and eventually falls on the gapped state. This would again leave a gapped state invariant or result in an unexpected emergent symmetry. So without introducing 
a gapped-state emergent symmetry, the path can only centre around the gapless state or the TQCP and then has to connect two topologically distinct states with protecting symmetry $G_p$.

This is only possible if a) $U_D({\lambda})$ transforms non-trivially under $G_p$ or non-invariant under $G_p$ and b) states traced out are without protecting symmetries so that the path ${\mathcal{C}_T}$ breaks the protecting symmetry $G_p$.
For instance, if a) were not true, $U_D(\lambda)$ leaves all states with protecting symmetry $G_p$ and hence all in the same topological phase as $|\Psi_{T0} \rangle$. This again is inconsistent with the observation of a smooth connection between two topologically distinct states by $U_D(\lambda)$ centred around a TQCP.

Such a path traced out by local unitary transformation can't cross a line of critical points of Wilson-Fisher type as all states along the path are gapped. a) and b) basically indicate that there shall be a smooth deformation 
between topological states through gapped states without protection symmetry $G_p$. And topological states are stable only when the protection symmetry $G_p$ is unbroken.

Moreover, emergent symmetries and corresponding transformations in the gapless state always imply an ultra-violet invariance of surrounding gapped states. All the local operators in these gapped states shall have the same scaling dimensions in the short distance and naturally belong to the same phase from the scaling points of view.  

This was also the starting point of the previous section. So now we have established an equivalence between smoothly deforming a symmetry protected topological state into a topologically distinct one when the protecting symmetry is broken or relaxed and 
an emergent {\em continuous} symmetry at a TQCP.

In the next two subsections, we will discuss why such emergent symmetries do not usually appear in the standard Landau paradigm of order-disorder transitions and generalize the emergent symmetry concept to symmetry protected gapless phases which are infrared stable, unlike a TQCP.

 \begin{figure}
\includegraphics[width=6cm]{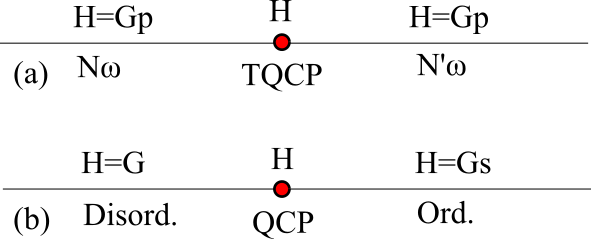}
\centering
\caption{ \footnotesize a) Symmetries in two gapped topological phases with different topological invariants $N_w, N'_w$ and at a typical TQCP where a toplogical phase transition occurs. The gapped phases are protected by symmetry $G_p$ which is also the symmetry of the ground states. $H=G_p \ltimes U_{EM}$ and the symmetry at the gapless TQCP is higher than $G_p$ because of emergent symmetries, i.e. $G_p \subset H$.  b) Near an ordered-disordered phase transition.
symmetries in disordered and ordered phases are $G$ and $G_S \subset G$ respectively. At a generic  gapless QCP, the symmetry group $H$ is equal to $G$, but not higher ($G$ shall also be the symmetry of interactions).}
\label{TQCPEMS1}
\end{figure}

\subsection{Distinction between TQCPs and QCPs in the standard Landau Paradigm}

In the standard Landau paradigm, a ground state can be invariant under group $H$. $H$ could be a subgroup of the symmetry group of the interactions $G$ or $H \subset G$.
The coset of $H$ is $M_g=G/H$  that defines the ground state manifold. In maximally disordered states, $H=G$, then $M_g$ has only one element and the ground state is non-degenerate.
However, in a maximumly ordered state, $H=I$ because of spontaneous symmetry breaking and its coset is just $G$ itself.

In many realistic applications, quantum critical fields are also invariant under $H=G$ just like in disordered states. So the gapless ground state manifold $M_g=G/G$ is non-degenerate represented by a Wilson-Fisher fixed point with space-time 
scale-conformal symmetry.  In the general cases, $M_g=G/H$ with $H \subset G$. Broadly speaking, the symmetry group at QCPs, $H$ is lower or at mostly equal to $G$. In situations of multi-criticality, $H$ can be enhanced beyond $G$ but those are not generic from the point of view of quantum criticality and we will not elaborate here.

In order to make a connection to TQCPs discussed in the previous section, we now ask what can happen if we lower the symmetry $G$ by adding symmetry breaking interactions. Let us assume that the symmetry group becomes $G_S \subset G$ where the subscript {\em S} implies a smaller symmetry group.
The question we are going to address is: Can we now smoothly and continuously deform a disordered non-critical state $|\Psi_0\rangle$ into an order one $| \Psi_1\rangle$ on the other side of  a QCP defined by the symmetry group $G$ via taking the initial state across the Hilbert space
with a lower symmetry $G_S$? 

If we restrict to our deformation to transformations out of local unitary operations, then generally this is forbidden by the principle of spontaneous symmetry breaking. Consider a global unitary transformation $U_D(\lambda)$ which is again generated by a local operator $\mathcal{O}({\bf R})$ via $\sum_{\bf R} \mathcal{O}({\bf R})$. For measurable physical operators $F({\bf r})$, we can define the following correlations,

\begin{eqnarray}
C_0 ({\bf r}_1 -{\bf r}_2) &= & \langle \Psi_0| F({\bf r}_1)  F({\bf r}_2)|\Psi_0\rangle,  \\
C_1 ({\bf r}_1 -{\bf r}_2) &=& \langle \Psi_1 | F({\bf r}_1) F({\bf r}_2) |\Psi_1 \rangle.
\end{eqnarray}

If there existed a unitary transformation so that $|\Psi_1\rangle =U(\lambda=1)| \Psi_0\rangle$, then $C_1$ has to be related to $C_0$ via

\begin{eqnarray}
C_1 ({\bf r}_1 -{\bf r}_2) &=& \tilde{C}_0 ({\bf r}_1 -{\bf r}_2)
=\langle \Psi_0| \tilde{F}({\bf r}_1)  \tilde{F}({\bf r}_2)|\Psi_0\rangle \nonumber \\
\tilde{F}({\bf r}) & =& U(\lambda=1) F({\bf r}) U^{-1}(\lambda=1)
\label{C01}
\end{eqnarray}
where one can show that $\tilde{F}({\bf r})$ is also a {\em local operator} if $U$ are local unitary transformations.
Furthermore, for the particular deformations studied below, the correlation length is also conserved. They can be further generalized to the ones without preserving correlations, similar to what had been proposed in Ref.\cite{Hastings05}.

Following the principle of the operator- product- expansion, in the UV limit when ${\bf r}_1-{\bf r}_2 \rightarrow 0$, all products of these local operators can be expanded linearly in terms of  a set of local operators of given scaling dimensions. 
This implies that

\begin{eqnarray}
C_1({\bf r} \rightarrow 0)=\tilde{C}_0({\bf r} \rightarrow 0) \sim C_0({\bf r} \rightarrow 0)  \sim \frac{1}{r^{2\eta_F-d_p}}
\label{CE}
\end{eqnarray}
where $\eta_F$ is defined as the scaling dimension of the local operator $F$ and $d_p$ is the scaling dimension of a local primary operator in the conformal field theory that dominates the ultraviolet dynamics.

 \begin{figure}
\includegraphics[width=6cm]{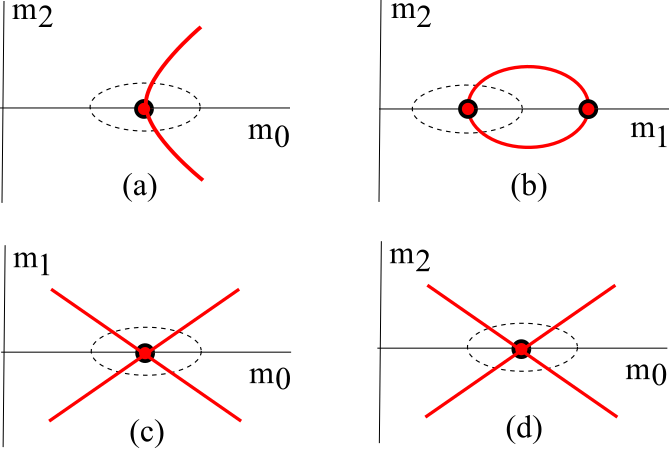}
\centering
\caption{ \footnotesize  A conventional order-disorder QCP (red dots) with Hamiltonian symmetry $G$ defined along the horizontal axis (either $m_0$ as in a), c) and d) or $m_1$ in b)). In the rest of the plane, the symmetry $G$ can be either {\em partially explicitly broken} or simply lowered to $G_S \subset G$.
a) A generic situation where a QCP along the axis of $m_0$ with symmetry $G$ is extended into an open line (red) of critical points into the $m_0$-$m_2$ plane when a subgroup symmetry in symmetry group $G$ is further explicitly broken. 
b) Another situation where QCPs with symmetry $G$ become a closed loop (red) of critical points in the plane when a subgroup symmetry of the Hamiltonian symmetry $G$ is broken beyond the horizontal symmetry axis of $m_1$.
In this case, there can be a second QCP along the symmetric axis. As illustrated in Appendix A, an example of b) is when the $Z^T_2$ subgroup in $G=Z_2 \otimes Z^T_2$ along the $m_1$-axis is further broken in the $m_1-m_2$ plane when moving away from the horizontal axis of $m_1$.
In c) and d) when the symmetry $G$ is simply lowered into a smaller group $G_S$ in either the $m_0$-$m_1$ or $m_0$-$m_2$ plane away from the horizontal axis of $m_0$, generically the QCP along the axis of $m_0$ with Hamiltonian symmetry $G$ becomes an intersecting point of two critical lines (See Appendix A for 
discussions on $G=U(1) \rtimes Z^T_2$ lowered into $G_S=Z_2 \otimes Z^T_2$). 
In a)-d), smooth deformations around a QCP (dashed oval enclosing the red dots) are strictly forbidden because of a line or lines of critical points emitted from the QCP located along the horizontal symmetry axis. }
\label{LandauPD}
\end{figure}

However, in the infrared limit, near a generic QCP, correlations of local operators such as $F({\bf r})$ on two sides of a QCP shall exhibit distinctly different properties because of the presence (absence) of long range orders in the ordered (disordered) phase. At large distances, $C_{0,1}$ are controlled by different infrared stable fixed points. One breaks the symmetry group $G$ down to $H \subset G$ and the other one is symmetric under the action of $G$. Eq.\ref{C01} on the other hand implies an equivalence of correlations in states $|\Psi_{0,1}\rangle$ as far as the long range order is concerned. And it can not be valid near a QCP.

The absence of such smooth deformations around a QCP that connect two states in two different phases of the symmetry group $G$ suggests that a disordered state can only be deformed into another disordered state, but not to a state in the ordered phase. 
In addition, paths traced out by the unitary transformations shall be bounded a critical line or a hyper-surface of Wilson-Fisher critical points where long ranged correlations appear. 
This in turn indicates that a gapless QCP under symmetry group $G$ shall be naturally extended into 
gapless critical states with lower symmetries such as $G_S$ so that deformations across QCPs are strictly forbidden even when the symmetry $G$ is relaxed.

In the Appendix A, we will illustrate this in a concrete model.  We show that no continuous symmetry such as $U_D$ symmetry shall emerge at a generic QCP if the symmetry group $G$ is either {\em partially broken} explicitly or lowered into a smaller symmetry group $G$. Here, we use {\em partially broken} to refer to breaking a specific subgroup symmetry.   
We examine two simple symmetry groups:
either $G=U(1) \rtimes Z^T_2$ or $G=Z_2 \otimes Z^T_2$.

The important conclusion is although universality classes of QCPs crucially depend on symmetry group
$G$ (as well as dimensionalities), presence of QCPs in the context of order-disordered transitions is very robust and persists even when a symmetry group $G$ is {\em partially broken} or lowered to $G_S$. Generically, QCPs shall form a line or hypersphere separating order-disordered phases when symmetry groups are varied. The universalities of QCPs along such a line are usually different depend on the symmetry group at a particular QCP. The robustness can be attributed to the general condensation phenomenon that shall always occur along the line and need no symmetry protection. This distinguishes the Landau paradigm from TQCPs discussed below.

Let us assume that a mass operator with symmetry $G$, say $m_0$ (or $m_1$), forms a one-dimensional parameter space and a QCP sits at $m_{c}$ along the symmetry axis defined by $m_0$ or $m_1$ (See Fig.\ref{LandauPD}).
When mass operators are further extended to include operators of non-zero $m_2$ that break a subgroup symmetry in $G$ or lower the symmetry $G$ to $G_S$, then there are a few possible scenarios of how a QCP with symmetry $G$ along the axis $m_0$
(or the axis $m_1$) extends into a line of QCPs in the plane of $m_0-m_2$ (or $m_1-m_2$ plane) etc. 

One very natural one is that a QCP simple extends into an open un-compact line of fixed points when subgroup symmetry is broken explicitly.
Second possibility is a QCP instead becomes a point on a compact line or a closed loop in the plane. We show such a case in a model when  the time reversal symmetry subgroup $Z^T_2$ in symmetry group $G=Z_2 \otimes Z^T_2$ is further broken
when additional mass operators $m_2$ are introduced.

It is also possible that when symmetry $G$ is broken  a QCP at a critical value $m_{c}$ with symmetry $G$ becomes a crossing point of two or more extended lines with lower symmetry $G_S$ (See Fig.\ref{LandauPD}). 
One such example given in the Appendix  A is  a model with $G=U(1) \rtimes Z^T_2$ and what happens when its symmetry is simply lowered to $G_S=Z_2 \otimes Z^T_2$. At such a QCP, the ground state symmetry is $H=G$
while the symmetry of the rest of the critical line is $H=G_S$.
 (When this happens, a QCP with symmetry $H=G$ can alternatively be viewed as a multiple critical point from the point view of lower symmetry QCPs with symmetry $G_S$.)

In all cases that interest us, continuous deformations around a gapless QCP with symmetry $G$ smoothly connecting two states on two sides of a QCP respectively are forbidden and there will be no emergent continuous symmetries unlike in TQCPs discussed in the next few sections.

\subsection{Symmetry protected gapless phases and adiabatic deformations without protecting symmetries}

Before concluding our general discussions, we briefly comment on deformations around a gapless phase with protecting symmetries. The discussions can be carried out parallel to what we have had on a TQCP except that
the gapless states now form a stable phase. We shall represent the phase by a finite segment along an axis of relevant mass operators with protecting symmetry $G_p$ rather than a point in the parameter space (See Fig.\ref{TQCPEMS}).
The parameter space that is perpendicular to the chosen axis of $G_p$ or simply the axis $G_p$ breaks the symmetry by including mass operators $M_1, M_2$ with lower symmetry $G_S \subset G_p$ and we call the plane $M_1$-$M_2$ perpendicular to 
the axis of $G_p$. 

In order for continuous symmetries to emerge in the gapless phase via the mechanism of deformations, we would have to allow that a gapped state with symmetry $G_S$ surrounding the gapless states along axis $G_p$ can be deformed into each other
via a closed loop. Such a loop again shall always be pierced by the axis $G_p$. A deformation path not pierced by $G_p$-axis sits entirely inside the gapped phase, away from the axis of gapless phases with symmetry $G_p$. Again, when contracted to the center of the path loop, the gapped state at the centre develops an emergent symmetry which shall not be happening as an emergent symmetry is not a lattice symmetry. So the path has to be non-contractable in the gapped phase and can only be contracted when it falls on the axis of $G_p$. A closed loop shrinks into a zero size one only when the deformed state is gapless.

On the other hand, the deformations induced by unitary transformations $U_D$ in the gapped phase around the axis $G_p$ always leave the gapless state invariant and becomes an emergent continuous symmetry in the gapless limit if the gapless phase is represented by an isolated fixed point.
Just like in the case of TQCPs, the continuous deformations in gapped states with lower symmetry $G_S$ around the axis $G_p$ indicates an emergent continuous symmetry in the gapless states along the axis of $G_p$.

\section{Summary of our main results}

Before we carry out detailed discussions in more concrete models, we first summarize our main findings. Two main questions we are going to address below are

i) why are there emergent symmetries at TQCPs or in symmetry protected gapless phases and how they are related to gapped phases around them?

ii) how general are these emergent symmetries and to what extent do they, either continuous or discrete ones, depend on fixed points involved, strong coupling or weakly interacting?

 \begin{figure}
\includegraphics[width=6cm]{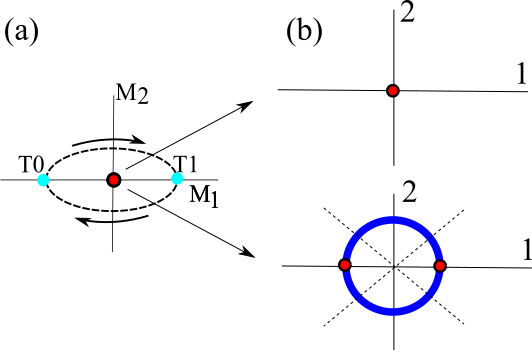}
\centering
\caption{ \footnotesize TQCP as an Isolated fixed point vs as a fixed point located in a smooth manifold of other fixed points without protecting symmetries.
a) A closed path (dashed oval) of deformations of gapped state around a TQCP (red dot) with protecting symmetry $G_p$ in the plane of $M_1$-$M_2$; the states along the axis of $M_1$ are symmetric under $G_p$ but the rest of the plane breaks the
protecting symmetry $G_p$. $T0,T1$ (blue dots) are two topologically distinct states with symmetry $G_p$. b) Two scenarios for a gapless TQCP (at $M_{1,2}=0$); the horizontal (1) and the vertical(2)-axis are, respectively, for interactions with and without protecting symmetry $G_p$. The top one is when the TQCP is an isolated infrared stable non-interacting fixed point that has the protecting symmetry $G_p$ and the gapless state is unique. Deformations leave it invariant. The bottom one is for the strongly interacting limit where there can be a smooth manifold of conformal field theory fixed points. Here a TQCP is identified with one of two fixed points (Red dots along the horizontal axis) with protection symmetry $G_p$. The deformations applied to a TQCP trace out all conformal fields along the ring-like manifold that break the protecting symmetry $G_p$ and hence do not leave the TQCP invariant. Two fixed points on the horizontal axis are dual to each other. Dashed lines here imply possible separatrices in renormalization flows towards the ring manifold. See section VII and Appendix.D for details.}
\label{TQCPEMS2}
\end{figure}

Our main conclusions are

A) Unlike at conventional QCPs in order-disordered transitions where $H$, the symmetry of QCP can be the same as the symmetry of interactions $G$, at TQCPs with protection symmetry $G_p$, the symmetry group can be $H=G_p \ltimes U_{EM}$.
The emergent symmetry group $U_{EM}$ is closely related to $U_D(\lambda)$, a group of deformations
around the gapless TQCP. If the TQCP with protection symmetry $G_p$ is represented by a well isolated scale-invariant fixed point (See Fig.\ref{TQCPEMS2}), 
one can show that $U_{EM}$ shall be precisely equal to $U_D(\lambda)$, the group of deformations of gapped phases around the TQCP that smoothly connects two topologically distinct gapped phases with protection symmetry $G_p$.

Such continuous deformations $U_D(\lambda)$  when applied to a gapped topological state can always trace out gapped states without protecting symmetry $G_p$ or with lower symmetries as $U_D(\lambda)$ breaks the protecting symmetry.
The only state with protecting symmetry $G_p$ that can be invariant under $U_D(\lambda)$ is the gapless state enclosed by deformed gapped states.

B) An obvious example of a well isolated scale invariant fixed point in the neighbourhood of a subspace protected by symmetry $G_p$ is an infrared stable non-interacting or weakly interacting fixed point.
The gapless state then is uniquely defined and deformations will always leave it invariant. We find an explicit form of deformations in a $3D$ quaternion superconductor where TQCPs with protecting symmetry $G_p=Z^T_2$ belong to the universality class of free real fermions.

C) The one-to-one correspondence between deformations and emergent continuous symmetries however breaks down when the TQCP with protecting symmetry $G_p$ is not represented by an isolated fixed point.
This can happen in strongly interacting limits where the fixed point of the TQCP with protecting symmetry $G_p$ locates itself in a manifold of other strong coupling fixed points without protecting symmetries.
In this case, in addition to deformations of a gapped topological state into gapped states without protecting symmetry $G_p$, $U_D(\lambda)$ also deforms the strong coupling fixed point of TQCP with symmetry $G_p$ into other fixed points with lower symmetries.
In fact, the manifold can be exactly traced out by the deformations $U_D(\lambda)$ and represents a set of conformal-field-theory fixed points without protecting symmetries except the ones related to TQCPs. 
In this case, the strong coupling TQCP with protecting symmetry $G_p$ is no longer symmetric under the deformation group $U_D(\lambda)$ and does not exhibit an emergent continuous symmetry.
It can nevertheless have a discrete $Z_2$ symmetry originating from a duality transformation preserving the protecting symmetry $G_p$.

 D) In general, continuous deformations of gapped states around a TQCP are a necessary but not a sufficient condition for emergent continuous symmetries.
 It becomes a sufficient one only when the TQCP is an isolated fixed point.

\section{Gapless states with emergent $U(1)$ symmetries}

So far we have argued a general relation between continuous deformations $U_D(\lambda)$ in gapped phases with symmetry $G_S (\subset G_p)$ and emergent symmetries $U_{EM}$ at TQCPs or other gapless states with protecting symmetry $G_p$.
We have illustrated that a gapless state can be thought as a singlet under the action of group $U_D(\lambda)$ if it is a well isolated fixed point while gapped states are usually $U_D(\lambda)$-charged and can be deformed by $U_D(\lambda)$.
Below we will examine these issues in a few concrete models.

\subsection{Symmetry protected TQCPs with emergent $U(1)$ symmetries}

We start with 3D topological superconductors with protecting time reversal symmetry $Z^T_2=G_p$. The emergent $U(1)$ symmetry at a  gaplessTQCP in this limit was pointed out before; the corresponding symmetry group $U_{EM}$ is 
generated by one of the six generators of $Spin(4)=SU(2) \otimes SU(2)$ group \cite{Zhou23}. In order to visualize an explicit connection to smooth deformations of gapped states surrounding it that can be employed to connect two topologically distinct states,
we first  examine the deformations of gapped ground states in the BCS representation. Later on we generalize it via an effective field theory to track the deformations of its vacuum which captures non-mean-field physics beyond the BCS theory.

The gapped states under our considerations are of the structure of $p+is$,

\begin{eqnarray}
&& \Delta({\bf k})= i\sigma_y [ {\bf \sigma} \cdot {\bf e}_{\bf k} n_p({\bf k})+ i n_s({\bf k}) ] \Delta_0(|{\bf k}|), \nonumber \\
 && n_p^2({\bf k})+n_s^2({\bf k})=1, n_{p,s} \in R. \nonumber \\
 && n_p({\bf k}) =\frac{\Delta_p |{\bf k}|}{\Delta_0(|{\bf k}|)}, n_s({\bf k}) =\frac{\Delta_s}{\Delta_0(|{\bf k}|)}
\label{quaternion}
\end{eqnarray}
where ${\bf e}_{\bf k}$ is a unit vector along the momentum ${\bf k}$.

As $n_{p,s} \in R$, states in Eq.\ref{quaternion} are precisely quaternion states. 
An outstanding property of a quaternion state is its pairing amplitude lives in a $S^3$ sphere. 
Each point of the sphere $S^3$ defined by $({\bf e}_{\bf k} n_p, n_s)$ represents a quaternion state, with $\Delta_{p,s}$ being the p-, s-wave pairing amplitudes respectively.
 $\Delta_0(|{\bf k}|)=\sqrt{\Delta^2_p |{\bf k}|^2 +\Delta_s^2}$ shall be finite everywhere in the momentum space for arbitrary ${\bf k}$ except when $n_s=0$.
When $n_s=0$, $\Delta({\bf k}=0)$ vanishes at ${\bf k}=0$.

When $n_s=0$, Eq.\ref{quaternion} describes a $Z^T$ invariant topological superconductor in the DIII class.
A finite $n_s$ breaks the protecting time reversal symmetry $G_p=Z^T_2$ and lowers it to $G_S=Z_1$, i.e. no symmetry. 
So a finite $n_s$ potentially allows a deformation from a gapped topological state into another distinct one or a trivial one. 
We define $|\Delta_0 n_p, \Delta_0 n_s, \mu\rangle_{bcs}$ as a quaternion BCS state and
$\Delta_0 n_p$, $\Delta_0 n_s$ and $\mu$ are $p$-wave pairing amplitude, $is$-wave pairing amplitude and the chemical potential $\mu$, respectively.

To simplify our presentation, we assume a strong pairing limit where $m \Delta^2_p \gg  \mu$  and the one-particle dispersion $\epsilon_{\bf k}$ is very flat so that can be neglected when $|{\bf k}| < \Lambda_{uv}=\Delta_p m$, $\Lambda_{uv}$ being the ultraviolet (UV) scale of our theory (However, this assumption can be easily relaxed.).

Now we introduce the following deformation operator 

\begin{eqnarray}
&&D= \frac{1}{2}\int d{\bf r} [ \Psi^T({\bf r}) i\sigma_y {\Psi({\bf r})}-\Psi^\dagger({\bf r})i \sigma_y {{\Psi^\dagger}^T({\bf r})}],\nonumber \\
&&U_D(\lambda)=\exp [i\pi \lambda D].\nonumber \\
\end{eqnarray}

Pairing at momentum $({\bf k} -{\bf k})$ in a quaternion state only involve five out of total sixteen possible Fock states.
There are i) empty state or spin singlet with no particles or $N=0$, $|1\rangle =|S=0, N=0 \rangle$, ii) spin singlet pairing states with $N=2$, $|2\rangle =|S=0, N=2\rangle$,
iii) spin triplet pairing states $|3\rangle$, $|4\rangle$, $|5\rangle$ $\in |S=1, N=2\rangle$. 

We can restrict the deformation in the subspace of $(|1\rangle, |2\rangle)$ by the following modification of $D$;

\begin{eqnarray}
D \rightarrow \mathcal{P}_{bcs} D \mathcal{P}_{bcs}, P_{bcs}=\prod_{\{{\bf k},-{\bf k}\}} [ |1><1|+|2><2| ]_{\{{\bf k},-{\bf k}\}}  \nonumber \\
\end{eqnarray}
where we have further listed the projection operator $\mathcal{P}_{bcs}$ to project away non-BCS element induced by the transformations or deformations.

These deformations of a quaternion state lead to  

\begin{eqnarray}
&& U_D(\lambda) |\Delta_0 n_p,  \Delta_0 n_s, \mu\rangle_{bcs} =| \mathcal{R}\Delta_0 n_p, \mathcal{R} \Delta_0 n_s, \mathcal{R}  \mu \rangle_{bcs} \nonumber \\
&& \mathcal{R}(\lambda)  \mu=\mu \cos\lambda \pi \mu +\Delta_0 n_s \sin\lambda \pi ,\nonumber \\
&& \mathcal{R} (\lambda) \Delta_0 n_s=\Delta_0  n_s \cos\lambda \pi-\mu \sin\lambda \pi \mu, \nonumber \\
&& \mathcal{R}(\lambda) \Delta_0 n_p =\Delta_0 n_p.
\label{DF}
\end{eqnarray}
Eq.\ref{DF} indicates how the quaternion state can be deformed under the deformation transformation $U_D$. 
First of all, when $n_s=0$ and the state can be identified as a topological one in the DIII class with protection symmetry $G_p =Z^T_2$, the deformations lift the state to the quaternion state manifold with non-zero $n_s$ while lowering the protecting symmetry $G_p=Z^T_2$ to $Z_1$.
It effectively generates a rotation in the $\mu-\Delta_0 n_s$ plane. Second, when $\mu=n_s=0$, the states under deformations remain to be in a sub-manifold of quaternion states with $n_s$ being zero and with protecting symmetry $G_p$. We identify this as the gapless limit.

Third, if we apply to the case $n_p=0$ and the state is a trivial $is$-wave superconductor, the deformations leave the state in the $n_p=0$ quaternion submanifold.
Finally, it is imperative that $n_p$ under such deformations remains fully invariant and symmetric. One can also verify that for $|\Psi_{T0}\rangle$ with a finite positive $\mu$ and $n_s=0$ so that it is a gapped topological state with protecting symmetry
$G_p=Z^T_2$, at arbitrary $\lambda$, the deformed state $U_D(\lambda) |\Psi_{T0} \rangle$ remains to be fully gapped in the quaternion manifold.

As discussed in the previous section, the topological distinct states we want to deform and smoothly connect are $|\Psi_{T0, T1}\rangle$. 
In the quaternion manifold, states with $G_p=Z^T_2$ in DIII class can be expressed as

\begin{eqnarray}
|\Psi_{T0}\rangle &=& |\Delta_0, 0, \mu\rangle_{bcs}, \nonumber \\
|\Psi_{T1}\rangle &=& |\Delta_0, 0,-\mu\rangle_{bcs}
\end{eqnarray}
where two states are $Z^T_2$ invariant as $n_s=0$ and are in a submanifold of quaternion states.

More specifically in terms of deformation $U_D(\lambda)$, following Eq.\ref{DF},
\begin{eqnarray}
|\Psi_{T\lambda} \rangle&=&U_D(\lambda) | \Psi_{T0} \rangle  \nonumber \\
&=& |\Delta_0, -\mu\sin\lambda \pi,  \mu \cos\lambda \pi \rangle_{bcs}.
\end{eqnarray}
In this case, among all the states along the path of $\lambda \in [0,1]$, only at $\lambda=0, 1$ states are $Z^T_2$ invariant and $n_s=0$ corresponding to $|\Psi_{T0,T1}\rangle$. The rest all break the $Z^T_2$ symmetry.

The continuous deformations of $U_D(\lambda \in [0,1])$ without crossing a critical line in the $\mu-\Delta_0n_s$ plane with lower symmetry $G_S=Z_1$ or no symmetry
therefore trace out fully gapped quaternion states between $\lambda=0$ and $\lambda=1$ without protecting symmetry $G_p$. The path $\mathcal{C}$ defined by $U_D(\lambda \in [0,1])$
smoothly connects two states $|\Psi_{T0,T1}\rangle$ via lowering the symmetry from $G_p$ to $G_S$ as anticipated in the previous general discussions.

If we choose the state to be at a gapless TQCP with $\mu=0$, then following Eq.17 the same deformations $U_D({\lambda})$ keep such a state in the submanifold where $n_s=\mu=0$ for arbitrary $\lambda$.
This thus clearly leave the gapless state of TQCP invariant hence leads to an emergent symmetry.
That is the smooth deformations in gapped phases without protecting symmetry $G_p=Z^T_2$ that connect two topologically distinct states are precisely the same emergent symmetry group that a gapless TQCP shall have.
Because of the emergent symmetry, the gapless states are invariant under $H=Z^T_2 \rtimes U_{EM}$, $U_{EM}=U_D(\lambda)$, instead of $G_p=Z^T_2$. Let us emphasize that in superfluids the charge $U(1)$ is broken and the deformation group $U_D(\lambda)$ is not associated with the conventional charge $U(1)$ symmetry.

\subsection{Symmetry protected Gapless phases with emergent symmetries}

Now we turn to another example of smooth deformations, a 3D gapless nodal phase with protecting symmetry $G_p=Z_2$ being the parity group\cite{Wen02,Beri10,Kobayashi14,Schnyder11,Matsuura13,Yang21,Armitage18}.
The state we are considering can be a $k_x+ik_z$ chiral state in two components $L$, $R$ or $\sigma_y=\pm1$. The locations of $L,R$ nodal points at $\sigma_y k_{y0}$ can be physically tuned by an external magnetic field.
Indeed, the gapless state can be conveniently created by applying a large external magnetic field along the $y$-direction to a time reversal symmetric topological superconductor.
 So all the continuous symmetries such as $SU(2)$ spin rotation symmetry, charge $U(1)$ symmetries  as well as the time reversal symmetry are broken because of underlying spin-orbit couplings, pair condensation and external fields respectively.
 The only symmetry survives in the phase is a discrete reflection symmetry $P$ (see below).

 The external magnetic field around the $y$ direction is further renormalized by pairing in the topological state and can have an effective form of $\sigma_y B^{eff}_z$, $B^{eff}_y= c k_y $, proportional to the $y$-component momentum $k_y$ measured from the $L, R$ nodal points $\pm k_{y0}$. 
Note again that the underlying spin-orbit coupling in topological superconductors breaks the spin rotation symmetry, either $SU(2)$ or $U(1)$ subgroup symmetry.
 
 However, a reflection through the plane of $y=0$, does leave the gapless state invariant as 
 \begin{eqnarray}
 k_x \rightarrow k_x, k_y \rightarrow -k_y, k_z \rightarrow k_z, L \rightarrow R,  \sigma_y \rightarrow - \sigma_y. \nonumber \\
 \label{reflection}
 \end{eqnarray}
The last equality together $k_y$ reflection result in a desired reflection symmetry of the effective pseudo magnetic field $B^{eff}_y$, i.e. $B^{eff}_y \rightarrow B^{eff}_y$.
 In fact, such gapless states are protected by the parity symmetry associated with the reflection $P$, and $P^2=1$.

The states are no longer gapless when the $\sigma_{x,z}$ couplings with the effective pseudo fields, $\sigma_{\alpha}B^{eff}_{\alpha}$, $\alpha=x,z$, are also present.
These two pseudo fields represent couplings between the two nodal points $L, R$ or $\sigma_y=\pm 1$ degrees of freedom in the nodal point phase and play the same role as a magnetic field to a physical spin.
Under the same reflections, $B^{eff}_{x,z} \rightarrow - B^{eff}_{x,z}$ and so their presence explicitly break the protecting parity symmetry $G_p$.
We can choose the following gapped BCS state near a chiral gapless phase in an effective field ${\bf B}^{eff}$,

\begin{eqnarray}
\Delta({\bf k})=\Delta_0 [ k_x+ik_z] \mathcal{I}_{2\times 2},  {\bf B}^{eff}=(B_0, B^{eff}_y=ck_y, 0) \nonumber \\
\end{eqnarray}    
where the gapless phase is stable when $B^{eff}_x=B^{eff}_z=0$ with the protecting parity symmetry $G_p=Z_2$.
With $B^{eff}_x=B_0 \neq 0$, the protecting symmetry $G_p$ is explcitly broken and states become fully gapped.

At $({\bf k}, -{\bf k})$ states, pairing in the nodal phases involve four different states: i) empty states; ii) two two-particle pairing states at $L$ or $R$ points;
iii) one four-particle states involving two pairs at both $L$ and $R$ nodal points.

We can introduce the deformation operator related to the following pseudo spin operator 

\begin{eqnarray}
D= \int d{\bf r} \Psi^\dagger({\bf r}) \sigma_y \Psi({\bf r}), U_D(\lambda) =\exp[i \lambda \pi D], \lambda \in [0,2]. \nonumber \\
\end{eqnarray}
When applied to a general gapped state specified as 

\begin{eqnarray}
|\Psi \rangle =|\Delta_0; (B^{eff}_x, B^{eff}_y, B^{eff}_z) \rangle, \nonumber \\
\end{eqnarray}
$U_D(\lambda)$ further deforms the state around the $B_y$ axis while leaving all deformed states fully gapped,

\begin{eqnarray}
U_D({\lambda}) |\Psi \rangle&=&|\mathcal{R}\Delta_0;(\mathcal{R} B^{eff}_x, \mathcal{R} B^{eff}_y, \mathcal{R} B^{eff}_z) \rangle \nonumber \\
\mathcal{R} \Delta_0&=&\Delta_0, \mathcal{R} B^{eff}_y = B^{eff}_y, \nonumber\\
\mathcal{R} B^{eff}_x &=& B^{eff}_x\cos\lambda \pi +B^{eff}_z \sin\lambda \pi, \nonumber \\
\mathcal{R} B^{eff}_z&=& B^{eff}_z\cos\lambda \pi-B^{eff}_x \sin\lambda \pi.
\end{eqnarray}

Now we consider an initial gapped state near a nodal point phase specified as 

\begin{eqnarray}
|\Psi_{0} \rangle =|\Delta_0; (B^{eff}_x=B_0, B^{eff}_y=c k_y, B^{eff}_z=0) \rangle, \nonumber \\
\end{eqnarray}
again $U_D(\lambda)$ further deforms the state around the $B_y$ axis and the closed path is simply defined by the circle ${B_x^{eff}}^2+{B^{eff}_z}^2=B_0^2$ (see Appendix.B for details).
$\lambda \pi$ conveniently defines the location of $(B^{eff}_x, B^{eff}_z)$ around the circle. In Fig.\ref{TQCPEMS}, we have shown a similar deformation in an $M_1-M_2$ plane that is equivalent to the $B^{eff}_x-B^{eff}_z$-plane discussed here.

The deformation path is pierced by the axis of $B^{eff}_x=B^{eff}_z=0$ and so when we take the limit of $B_0=0$, the deformation acting on the gapless nodal phase state with arbitrary $L, R$ locations leave every state in the phase invariant.
The gapless phase therefore has a single emergent continuous symmetry group defined by the deformation group $U_D(\lambda)$. 

Note that because the charge $U(1)$ symmetry is broken in a superconducting phase, charge is not conserved
and  the deformation symmetry discussed here shall not be confused with the conventional $U(1)$ symmetry. Furthermore, time reversal topological superconductors does not have either $SU(2)$ or $U(1)$ spin rotation symmetries so after applying an external magnetic field, both spin rotation symmetries and time reversal symmetry are absent and only a reflection $Z_2$ symmetry survives.

The deformation above $U_D(\lambda)$ transforms a gapped state non-trivially and does not leave a gapped state invariant. 
The protecting symmetry group $G_p=Z_2$ is discrete and there are no continuous symmetries in interactions and so not surprisingly, $U_D(\lambda)$ can {\em not} be the symmetry group of a gapped phase.
However, as seen here there can be an emergent symmetry group in the gapless phase with protecting symmetry $G_p=Z_2$ so the ground state symmetry group $H=G_p \ltimes U_{EM}$, $U_{EM}=U_D(\lambda)$ higher than the protection symmetry $G_p=Z_2$ itself.
The symmetry protected gapless phase here can be shown to be a {\em $\tau-\sigma$} dual of the TQCP discussed in the previous subsection\cite{Zhou23}.

This concludes our discussions on deformations of gapped states with lower symmetries and how these are related to emergent symmetries in gapless states with protecting symmetries $G_p$.
It is not entirely surprising that this symmetry protected gapless phase also exhibits an emergent infrared symmetry $U_{EM}=U(1)$ given the $\tau-\sigma$ duality between the nodal phase and a TQCP as pointed out in Ref.\cite{Zhou23}. 
In the next section, we turn to the EFT for TQCPs and study the vacuum of the EFT when acted on by the deformation operator $D$.  

\section{Emergent IR symmetries and Anomalies in EFTs}

To study the emergent IR symmetries beyond the mean-field physics studied in the previous section, we employ an effective field theory (EFT) the vacuum structure of which contains quantum fluctuations and correlations including conformal field theory physics well beyond the simple BCS theory.  
The effective field theory or EFT for gapless phases or TQCPs in superconductors was put forward previously. To discuss the deformations in gapped phases around them, we further generalize the EFT by including mass operators explicitly.
Without loosing generality, we can cast an interacting Hamiltonian of {\em massive real fermions} in the following infrared form,

\begin{eqnarray}
H_{eff} &=& H_0+\sum_{m=1,2} M_{m} \chi^T \beta_m\chi  \nonumber \\
H_0 &=&
 \frac{1}{2} \int d{\bf r} [ \chi^T({\bf r}){\bf  \alpha} \cdot {i\nabla} \chi ({\bf r})
\nonumber \\
&+& g_1 \chi^T \beta_1 \chi \chi^T  \beta_1 \chi +g_2 \chi^T  \beta_2 \chi  \chi^T \beta_2 \chi +... ] \nonumber \\
\label{Eff}
\end{eqnarray}
where ${\bf \alpha}=(\alpha_x,\alpha_y,\alpha_z)$. And $\alpha_{i}$, $i=x,y,z$ and $\beta_{1,2}$ are mutually anti-commuting Hermitian matrices.
That is,

\begin{eqnarray}
&& \{ \alpha_i, \alpha_j \}=2\delta_{i,j}, \{\beta_m,\beta_n \}=2\delta_{m,n}, \{ \alpha_i,\beta_m \}=0; \nonumber \\
&& \alpha^\dagger_i=\alpha_i, \beta^\dagger_m=\beta_m, i=x,y,z; m=1,2.
\label{Algebra}
\end{eqnarray}

We will restrict to real fermions with four components or $N_f=\frac{1}{2}$ in terms of four component Dirac fermions.
$N_f=\frac{1}{2}$ turns out to be a minimum number of degrees of freedom for our discussions on the specific protecting symmetries, or it forms a fundamental representation.

\begin{eqnarray}
&& \chi^T=(\chi_1,\chi_2,\chi_3,\chi_4), \chi^\dagger_i({\bf r}) =\chi_i ({\bf r}),  \{ \chi_i, \chi_j \}=\delta_{ij}. \nonumber \\
\end{eqnarray}
We have also chosen to introduce two interactions, $g_{1,2}$ for later discussions on emergent symmetries, although for the four component real fermions, these two operators turn out to be always equivalent.  
 And $\alpha$ matrices are real and symmetric ones while $\beta_{1,2}$ are antisymmetric purely imaginary matrices. That is,

\begin{eqnarray}
&& \alpha_i^T=\alpha_i=\alpha_i^*, \beta^T_m=-\beta_m=\beta_m^*; i=x,y,z, m=1,2. \nonumber \\
\label{Algebra2}
\end{eqnarray}

This EFT has an isolated free fermion fixed point at

\begin{eqnarray}
M_1=M_2=0, g_1=g_2=0.
\end{eqnarray}
This isolated scale invariant fixed point serves our purpose of discussions on emergent symmetries at TQCPs and continuous deformations of gapped states.

We will focus on weakly interacting TQCPs around this isolated free-particle fixed point in this Section. However, the EFT does have a strong coupling fixed point as $g_1 \sim g^*=1-d$ in $d=2,3$.
However, this strong coupling limit physics is much more convenient to study using its ultraviolet completion of the four-fermion interaction theory. Details are presented in Section VII. The physics near the strong coupling fixed point turns out to be very 
different from the ones discussed below because of strongly interacting emergent bosons and a manifold of fixed points with lower symmetries. We will come back in Section VII.

In our representations, we further introduce a $SU(2)$ subgroup of an Spin(4) group where emergent symmetry groups can be easily studied.
The three generators, $q_i, i=1,2,3$ can be related to two mass generators $\beta_{1,2}$ in a simple way,

\begin{eqnarray}
q_1&=&\gamma=i\alpha_x\alpha_y\alpha_z,
q_2= \beta_1=\beta_0, 
q_3=\beta_2=i \beta_0 \gamma,  \nonumber \\
q_i&=&q_i^\dagger=-q_i^T=-q^*_i, \{\beta_0, \alpha_i \}=\{\beta_0, \gamma\}=0.
\label{su2}
\end{eqnarray}
One can verify the above algebra straightforwardly by taking into account Eq.\ref{Algebra},\ref{Algebra2} and anti-communting relation between $\beta_0$ and $\alpha_i$, $i=1,2,3$.
Furthermore, $q_i, i=1,2,3$ that are antisymmetric pure imaginary hermitian generators indeed form an $su(2)$ algebraic group.

 \subsection{TQCPs with $G_p=Z^T_2$}

Now for discussions on symmetry protected TQCPs under a global protecting symmetry $G_p=Z^T_2$, i.e. time reversal symmetry, we can further enforce that $\beta_0=\beta_1$ has an even parity under the time reversal transformation and $\alpha_i$ $i=1,2,3$ have to be odd so that the Hamiltonian is time reversal invariant (TRI) 
under the time reversal transformation. That is,

\begin{eqnarray}
\mathcal{T}^{-1} \alpha_i \mathcal{T}=-\alpha_i, \mathcal{T}^{-1} \beta_{0} \mathcal{T}=\beta_0.
\label{TRI1}
\end{eqnarray}

Eq.\ref{TRI1},\ref{su2} further indicate that

\begin{eqnarray}
\mathcal{T}^{-1} \gamma \mathcal{T}= \gamma, \mathcal{T}^{-1} \beta_2 \mathcal{T}=- \beta_2.
\label{TRI}
\end{eqnarray}
That is the mass operator $M_2$ associated with $\beta_2$ defined above always breaks the time reversal symmetry while $M_1$ is time reversal invariant. 

As we have seen before, to study continuous deformations of a gapped phase, it is imperative to lift the Hamiltonian off the subspace protected by $G_p$ so to create a manifold of gapped states with lower symmetries $G_S$.
By allowing $M_2$ terms in the EFT, we are able to continuously deform the gapped vacuum of the EFT.
The gapped vacuum describes the ground state which in principle captures the non-BCS effects when interactions $g_{1,2}$ are present. So at least in the limit that EFT is applicable, the deformations of EFT can be applied to a broader class of states with richer structures than the simple BCS states we have used to illustrate the main ideas.

Once we identify the vacuum of our EFT as a ground state in topological calsses, instead of deforming the vacuum directly, we deform the Hamiltonian and infer the property of the deformed vacuum via the following relations.
\begin{eqnarray}
H_{eff} (M_1, M_2) |vac\rangle_{[M_1, M_2]} &=&E_{g} |vac\rangle_{[M_1, M_2]} \rightarrow \nonumber \\
\tilde{H}_{eff}(M_1, M_2) |\tilde{vac}\rangle_{[M_1, M_2]} &=& E_{g}|\tilde{vac}\rangle_{[M_1, M_2]}.
\end{eqnarray} 
Here {\it tilde} refers to the deformed EFT or vacuum under the deformation of $U_D(\lambda)$,

\begin{eqnarray}
\tilde{H}_{eff} (M_1, M_2) &=&U_D(\lambda) H_{eff} (M_1, M_2) U_D^{-1}(\lambda), \nonumber \\
| \tilde{vac}\rangle_{[M_1, M_2]} &=& U_D(\lambda) |vac\rangle_{[M_1, M_2]}.
\end{eqnarray}

Now if we define the deformations in terms of 

\begin{eqnarray}
D=\frac{1}{2}\int d{\bf r} \chi^T ({\bf r}) q_1 \chi ({\bf r}), U_D(\lambda)=\exp(i\frac{1}{2} \lambda \pi D)
\label{DO}
\end{eqnarray}
where $q_1$ together with $q_{2,3}$ or $\beta_{1,2}$ form an $su(2)$ group algebra.
One can easily show that

\begin{eqnarray}
&&\tilde{H}_{eff}(M_1, M_2) = H_{eff} (\tilde{M}_1,\tilde{M}_2) \nonumber \\
&&\tilde{M}_1  =\cos\lambda\pi M_1 -\sin\lambda\pi M_2, \nonumber \\
&&\tilde{M}_2 = \cos\lambda\pi M_2 +\sin\lambda\pi M_1.
\end{eqnarray}
 To obtain this result, we have utilize the result that $D$ commutes with $H_0$ defined in Eq.\ref{Eff} and the EFT with interactions $g_{1,2}$ or $H_0$ remains symmetric under the same $U_D(\lambda)$.
 
 We now can assign the vacuum of $H(M_1, 0)$ as the initial topological state with protection symmetry $G_p=Z^T_2$ that we want to deform.
 By applying $U_D(\lambda=1)$ to $|\Psi_{T0}\rangle$, we obtain the final gapped state $|\Psi_{T1} \rangle$ with the same protection symmetry $G_p$. But they are topologically distinct.
 However, under $U_D(\lambda)$, $\lambda \in (0,1)$, all deformed gapped states break the protecting symmetry $G_p$ because of finite $M_2$ as anticipated.  To summarize, we have
 
 \begin{eqnarray}
 |\Psi_{T0} \rangle=|vac \rangle_{[M_1,0]}, \Psi_{T1} \rangle =|vac\rangle_{[-M_1,0]}.
 \end{eqnarray}
 
 Note under the time reversal transformation $\mathcal{T}$,
 
 \begin{eqnarray}
U_{D*}(\lambda)= \mathcal{T}^{-1} U_D (\lambda) \mathcal{T} =U_D(-\lambda)=U_D(2-\lambda)   
 \end{eqnarray}
 so that $U_D(2-\lambda)$, $\lambda \in [0,1]$ can be applied to define $\mathcal{C}_*$ while $U_D(\lambda)$ defines $\mathcal{C}$ within the same domain of $\lambda$. Therefore, the closed path $\mathcal{C}_T=\mathcal{C}+\mathcal{C}_*$ is given by 
 $U_D(\lambda)$, $\lambda \in [0,2]$.
 
When applying $U_D(\lambda)$ to the gapless limit of $M_1=M_2=0$, the closed loop falls on the gapless state and $H_{eff}(0,0)$ is therefore invariant. This in turn verifies that the ground state or the vacuum is symmetric, or has an emergent symmetry exactly defined by the deformation operator $D$. One can show that in both the non-interacting limit and weakly interacting limit with finite but small $g_1$, the EFT always exhibits such an emergent $U(1)$ symmetry.

\subsection{Gapless phase with $G_p=Z_2$ being the parity group}

The same EFT can be applied to study the gapless nodal phase with protecting parity symmetry $G_p=Z_2$.
For that, we have to work with $\alpha_i$ and $\beta_0$ transformed properly under the reflection parity group $P$. Under the reflection through $y=0$-plane, we require that Eq.\ref{reflection} holds.
To maintain the invariance of the EFT under such parity transformations, we further introduce $\mathcal{P}$ as a unitary transformation defined in terms of anti-commuting $\alpha_i$ matrices,

\begin{eqnarray}
\mathcal{P}=i\alpha_x\alpha_y\alpha_z.
\end{eqnarray}

Therefore,

\begin{eqnarray}
\mathcal{P}^{-1} \alpha_{x,y,z} \mathcal{P}=\alpha_{x,y,z},
\label{PS}
\end{eqnarray}

Furthermore,

\begin{eqnarray}
\mathcal{P}^{-1} \beta_{1,2} \mathcal{P}=-\beta_{1,2},
\end{eqnarray}
where $\beta_{1,2}$ are both parity odd ensured by the relation $\beta_2=i \beta_1 \gamma$ (defined in Eq.\ref{su2}).
The EFT set up this way forms a representation of the gapless nodal phase with the protecting symmetry $G_p=Z_2$ when $M_{i}\beta_{i}$, $i=1,2$ are both set to be zero.
In  the fully gapped phases with finite $M_{1,2}$, the symmetry is lowered from $G_p=Z_2$ to $Z_1$.

The deformations of gapped phase can be done using the same operator in Eq.\ref{DO} as $q_1$ is part of the $su(2)$ algebra in Eq.\ref{su2} and generates the desired deformations around the gapless state with $M_{1}=M_{2}=0$.
The deformations lead a close loop defined by $U_D(\lambda)$, $\lambda \in [0,2]$ pierced by the gapless phase and become the invariant group of the gapless phase when the deformation loop falls on the gapless state(See Fig.\ref{TQCPEMS}). 
That ultimately again results in an emergent continuous symmetry in the gapless phase

In the following section, we will show that an EFT for a lattice model with the same symmetry group defined by $D$, or with lattice symmetry defined by $U_D(\lambda)$, is distinct from the EFT with such an emergent symmetry.
The EFT above can not be realized in lattices with the same symmetry in the same dimension . The EFT for TQCPs above therefore is anomalous from the point of view of the symmetry group $U_{EM}=U_D(\lambda)$.

\subsection{Emergent symmetries vs Lattice symmetries: Lattice models invariant under Emergent Symmetry Group}

The gapless state above is invariant under $H=G_p \ltimes U(1)$  due to emergent symmetries while the protecting symmetry $G_p=Z^T_2$ or $Z_2$ is lower than $H$.
Now we are going to examine a lattice theory with symmetry defined by $H$. We shall illustrate explicitly that the number of fermion degree of freedom $N_f$ for the symmetry group $H$ in a lattice model is always at least doubled compared to 
$N_f$ in the EFT with protection symmetry $G_p$. It is closely related the well-known issue of fermion doubling in lattice gauge theories and here manifests explicitly in gapless TQCPs or phases with protecting symmetry $G_p$.

The model in a three dimensional lattice with $H=G_p \ltimes U_{EM}$-symmetry, $G_p=Z^T_2$ and $U_{EM}=U(1)$ can be constructed as a subclass of the following generalized model:

\begin{eqnarray}
H&=&H_0 +H_M; \nonumber \\
H_0&=&\sum_{{\bf n}, {\bf a}} \chi^T_{\bf n} i {\bf \alpha}\cdot {\bf a} \chi_{{\bf n} +{\bf a}} +g_0\sum_{\bf n} \chi^T_{\bf n} \beta_0\chi_{\bf n}  \chi^T_{\bf n} \beta_0\chi_{\bf n}, \nonumber \\
H_M&=&\sum_{{\bf n}, {\bf a}} \chi^T_{\bf n} [ M_1 \beta_1 +M_2 \beta_2 ] \chi_{{\bf n} +{\bf a}}, \nonumber \\
\label{3Dlattice}
\end{eqnarray}
Here ${\bf n}=(n_x, n_y, n_z)$ with integer $n_i$, $i=x,y,z$ are lattice site indices and lattice constants are assumed to be unity.
${\bf \alpha}=(\alpha_x, \alpha_y,\alpha_z)$ are real symmetric vector matrices and $\beta_{1,2}$ antisymmetric ones as defined in the previous section.
${\bf a}=\pm {\bf e}_i$, $i=x,y,z$ are unit vectors pointing to nearest neighbouring sites along $x,y,z$ direction respectively. 

Imposing a lattice symmetry including the emergent $U(1)$ symmetry forces $M_{1,2}$ to be zero as finite $M_{1,2}$ breaks the emergent symmetry $U_{EM}=U_D$ that in our case is a $U(1)$ group.
The rest terms are invariant under the group of $H$. However, this lattice model leads to $2^3$ copies of gapless fermions in our EFT and they form a representation  of $N_f=\frac{1}{2}\times 8=4$ Dirac fermions 
rather than $N_f=\frac{1}{2}$. This is the fermion doubling problem generic to lattices with $U(1)$ symmetries such as the one under consideration and is independent of the lattice structure we have chosen to work with\cite{Nielson81,Friedan82,Kaplan92}.
Typically, such a theory has well-known t'Hooft anomalies which we refer to standard discussions in field-theory textbooks. So the $d$-dimension gapless states of $N_f=\frac{1}{2}$ fermions with both emergent symmetry and protection symmetry can not be realized 
in a $d$-dimension lattice of the same number of fermions $N_f$ and with the lattice symmetry $H=G_p\ltimes U_{EM}$ including both the protecting one $G_p$ and emergent symmetry $U_{EM}$.

In the next subsection, we however show that the EFT can emerge on a surface of $(d+1)$-dimension lattice model of $N_f=1$ fermions so to implement the same symmetry group $H=G_p \ltimes U(1)$ where both the protecting symmetry and $U(1)$ are subgroups.

\subsection{Constructing $(d+1)$-dimension theories with higher symmetries from $d$-dimension theory via Quantum doubling: $N_f \rightarrow 2N_f$}

We now show that a gapless TQCP or phase with a protecting symmetry $G_p$ and emergent $U(1)$ symmetry is naturally represented by a surface.
To construct a $d$-dimension surface theory with $H$ symmetry explicitly, we first generalize a $d$-dimension lattice model with protecting symmetry $G_p$ only to the $(d+1)$-dimension lattice theory with $H=G_p \ltimes U_{EM}$, $U_{EM}=U_D$.
So $d$-dimension theory and $(d+1)$-dimension theory have different lattice or {\em UV} symmetries and the difference is the emergent symmetry $U_D(\lambda)$. To generate the desired $(d+1)$-dimension lattice model with the {\em lattice} symmetry of $H$, higher than the symmetry group $G_p$ in the $d$-dimension theory we start with, we first double the number of fermions from $N_f$ in the $d$-dimension lattice, to $2N_f$ in the $(d+1)$-dimension lattice. This is sufficient to guarantee the $(d+1)$-dimension lattice model has a higher ultraviolet symmetry. Also in the process of elevating a $d$-dimension model to a $(d+1)$-dimension model, both before and after, the lattice models can describe both gapped symmetry-protected states and gapless TQCPs, but with different protecting symmetries. To summarize, we have

\begin{table}
\begin{center}
\begin{tabular}{|c|c|c|c|c|}
\hline
Dim. & $N_f$ & Latt. Sym.& Emerg. Sym. & Surf. Stat. \\
\hline \hline
3D &  $\frac{1}{2}$ & $G_p$ & $U_{EM}=U_D(\lambda)$ & Helical  \\
\hline
4D & 1 & $G_p \ltimes U_{EM}$ & $U^{(d+1)}_{EM}=\tilde{U}_{D}(\lambda)$ & Non-Chiral. \\
\hline
\end{tabular}
\caption{\label{table1} Elevated $4D$ model with a higher symmetry}
\end{center}
\end{table}

The above emergent symmetry structure in the gapless states in the $(d+1)$-dimension lattice model, $U^{(d+1)}_{EM}$, or the full deformation group $\tilde{U}_{D}(\lambda)$ will be further introduced later on.
 
Details of such constructions are presented in Appendix D for readers who are interested in a surface representation of a TQCP.
There, we explicitly construct a $(d+1)$-dimension lattice model, the surface of which can realize an EFT for $d$-dimensional TQCPs with both $U(1)$ emergent symmetries and protecting symmetry $G_p=Z^T_2$.
The $(d+1)$-dimension lattice turns out to be dual to a $4D$ {\em strong topological insulator} with charge $U(1)$ symmetry. An equivalence is established below between a $3D$ TQCP with $N_f=\frac{1}{2}$ gapless fermions in a topological superconductor without charge $U(1)$ symmetry, and a $3D$ surface of a $4D$ strong topological insulator with a charge $U(1)$ symmetry of $N_f=1$ fermions. Main differences between the construction of effective topological surfaces here for TQCPs and the domain-wall fermions in the standard quantum field  theories are also commented on in Appendix D.

\section{Emergent symmetries in the presence of strong interactions: Emergent bosons and Conformal fields}

So far we have focused on the limit of weak four-fermion interactions where the emergent symmetries at TQCPs can be conveniently related to infrared stable (with respect to interactions) weakly interacting fixed points and are robust as far as the low energy sector is concerned.
What happen to emergent symmetries when TQCPs are represented by strong coupling fixed points or conformal field theories (CFTs) are not very well understood, especially from the point of many-body state deformations. However for the purpose of studying deformations of gapless states and symmetries  near a strong coupling fixed point, the EFT with four-fermion interactions turns out to be inadequate.

Here we instead will illustrate the complexity of deformations in the gapless limit and dynamics
using an extended model involving real fermions interacting with one single {\it massless} scalar field. The emergent bosons that mediate long range interactions among fermions naturally result in strongly interacting fixed points that we can associate with a strongly
interacting TQCP well beyond the weakly interacting model so far we have dealt with. It is important to note that in the most generic situation of strongly interacting real fermions, we can couple fermions with only one single scalar field that is invariant under the global protecting symmetry group $G_p=Z^T_2$. As shown below, both emergent bosons and real fermions can be strongly interacting.

For our purpose, before projecting out the states with protecting symmetry $G_p$, we consider the following generalized model

\begin{eqnarray}
&&H_{eff} (m_0;g_{Y1}, g_{Y2}, g_{4}) = H_{0f} +H_{0b}+H_I, \nonumber \\
&& H_{0f} =  \int d{\bf r} [  \chi^T {\bf \alpha}\cdot i {\nabla}
\chi ({\bf r}) + \sum_{m=1,2} M_{m} \chi^T \beta_m\chi ], \nonumber \\
&&H_{0b} =\int d{\bf r} [ \pi^2({\bf r})  + \nabla \phi ({\bf r}) \cdot \nabla \phi({\bf r})  + m_0^2 \phi^2({\bf r}) ]; \nonumber \\
&& H_I = \int d{\bf r} [\sum_{m=1,2} g_{Ym} \phi({\bf r}) \chi^T({\bf r}) \beta_m \chi({\bf r}) +\frac{g_4}{4} \phi^4({\bf r})];  \nonumber \\ 
& & [ \phi({\bf r}) ,  \pi({\bf r}') ]  = i\delta({\bf r}-{\bf r'}), [\phi({\bf r}),\phi({\bf r'}]=[\pi({\bf r}),\pi({\bf r}')]=0. \nonumber \\
\label{CFT1}
\end{eqnarray}
where

\begin{eqnarray}
\chi^T= (\chi_1, \chi_2), \{ \chi_i({\bf r}), \chi_j({\bf r}') \} =\delta_{i,j} \delta({\bf r} -{\bf r}'), i=1,2. \nonumber \\
\end{eqnarray}
We denote the boson mass as $m_0$ and fermion masses as $M_{1,2}$.

Under the time reversal transformation ${\mathcal T}$, these real fields transform accordingly,

\begin{eqnarray}
\chi \rightarrow  \chi'=i\beta_2 K  \chi, \phi  \rightarrow \phi'=\phi, \pi \rightarrow  \pi'= \pi
 \label{TRS}
\end{eqnarray}
where $K$ is an action of complex conjugate and $\beta_2=i \beta_1 \gamma$, $\gamma=i\alpha_x\alpha_y\alpha_z$. Note that $\beta_1=\beta_0$ is even under the time reversal transformation while
$\beta_2, \alpha_i$, $i=x,y,z$ are odd. In addition, as stated before, $\beta_{1,2}$ are anti-symmetric hermitian and $\alpha_i$, $i=x,y,z$ are symmetric ones. 
To study gapped topological states with the desired protection symmetry $G_p$, in Eq.\ref{CFT1} we have to further demand $M_2=g_{Y2}=0$ as these operators explicitly break the protecting symmetry.
And the TQCP is identified with the limit when the time-reversal-invariant mass operator $M_1$ becomes zero.

\subsection{Fixed points and Fixed-point manifold in the strong coupling limit}

When $m_0$ is finite and bosons are massive, one can safely integrate out the gapped bosons and in the low energy sector this procedure leads to the four-fermion interactions that had been considered in previous sections.
As we will see below, this is exactly the limit when an emergent $U(1)$ symmetry appears in the gapless limit of $M_1=M_2=0$, i.e. $H=Z^T_2 \times U(1)$ as discussed extensively in the previous sections.   

Indeed for the scale much lower than the mass scale $m_0$, dynamics of scalar fields are locally independent. In this local-field-dynamic limit, it is convenient to set the scaling dimension of scalar fields as 

\begin{eqnarray}
[\phi]=L^{-\frac{(d+1)}{2}}, [\chi]=L^{-\frac{d}{2}},
\end{eqnarray}
and treat the rest kinetic terms as irrelevant operators.

Not surprisingly, a direct scaling analysis leads to the following equations in the tree level for coupling constants $g_{Ym}, m=1,2$ and $g_4$.

\begin{eqnarray}
\frac{d\tilde{g}_{Y1}}{d t} &=& \frac{d-1}{2}\tilde{g}_{Y1} +..., \nonumber \\
\frac{d\tilde{g}_{Y2}}{d t} &=&\frac{d-1}{2}\tilde{g}_{Y2} +..., \nonumber \\
\frac{d\tilde{g}_{ 4}}{d t} &=&\frac{d+1}{2} \tilde{g}_4 +...
\end{eqnarray}
where $t=\ln\Lambda$, $\Lambda$ is the running scale of the EFT. $\tilde{g}_{Ym}=g_{Ym}\Lambda^{\frac{d-1}{2}}$, $m=1,2$ and $\tilde{g}_4=g_4\Lambda^{\frac{d+1}{2}}$.  
It indeed explicitly indicates that in two-, three-dimensions, there is a well-isolated infrared stable fixed point 

\begin{eqnarray}
g_{Y1}=g_{Y2}=g_4=0,
\end{eqnarray}
i.e. an infrared stable free-particle fixed point which we can identify as a quantum critical theory.  A TQCP identified with this EFT with massive scalar field therefore shall belong to the universality class of free fermions.
We recover the weakly interacting TQCP physics discussed in the previous sections via a four-fermion interacting EFT. 
This infrared stable fixed point does have an emergent $U(1)$ symmetry because it is a fully isolated free-particle fixed point.

Below we will exclusively focus on the massless limit of $M_0=0$ so that real fermions at TQCPs can be strongly interacting due to their couplings to the emergent massless bosons.
This effectively represents an ultraviolet completion of the four-fermion interaction model where the strong coupling fixed point leads to the Gross-Neveu mechanism of mass generation\cite{Gross74}.  
In the massless limit, the scaling dimensions of the scalar field, together with those of real fermion fields, take a more standard form

\begin{eqnarray}
[\phi]=L^{-\frac{(d-1)}{2}}, [\chi]=L^{-\frac{d}{2}}.
\end{eqnarray}

 \begin{figure}
\includegraphics[width=6cm]{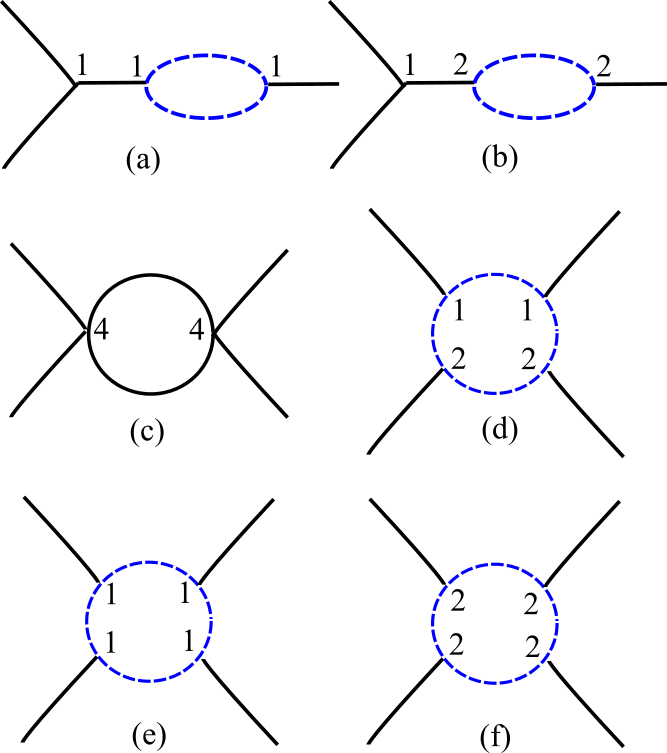}
\centering
\caption{ \footnotesize Examples of diagrams contribute to Eq.\ref{RGE} and the ring of fixed points in Eq.\ref{ring}.
Dashed (blue) lines are for real fermion fields and the solid  (black) ones for scalar fields.
$1,2$ and $4$ numerate the type interaction vertices appearing in our calculations. $1,2$ are for $g_{Y1,Y2}$ respectively while $4$ is for $g_4$. 
}
\label{RGED}
\end{figure}

The EFT in Eq.\ref{CFT1} in this limit allows strong coupling conformal field solutions or CFT states as its infrared stable fixed points. When the gap closes and $M_1=M_2=M_0=0$, the one-loop renormalization group equations for $H(0; g_{Y1}, g_{Y2}, g_4)$ derived in a diagrammatic approach  (See Fig.\ref{RGED}) have the following form,

\begin{eqnarray}
\frac{d\tilde{g}_{Y1}}{d t} &=& \frac{d-3}{2}\tilde{g}_{Y1} +\tilde{g}_{Y1} [\tilde{g}^2_{Y1} +\tilde{g}^2_{Y2}], \nonumber \\
\frac{d\tilde{g}_{Y2}}{d t} &=&\frac{d-3}{2}\tilde{g}_{Y2} +\tilde{g}_{Y2} [\tilde{g}^2_{Y1} +\tilde{g}^2_{Y2}], \nonumber \\
\frac{d\tilde{g}_4}{d t} &=&(d-3) \tilde{g}_4 +\tilde{g}^2_4 -c [\tilde{g}^2_{Y1} +\tilde{g}^2_{Y2}]^2. 
\label{RGE}
\end{eqnarray}
Here $\tilde{g}_{Yi}=g_{Yi}\Lambda^{\frac{d-3}{2}}$, $i=1,2$. $\tilde{g}_4={g}_4\Lambda^{d-3}$ are the dimensionless coupling constants and we find that $c=4$.

A very important feature here is that for dimensions lower than three or $d <3$, there are a ring of scale invariant fixed points that define a family of conformal field theories. The ring is defined as

\begin{eqnarray}
\tilde{g}^2_{Y1}+\tilde{g}^2_{Y2}=[\tilde{g}^*_Y]^2=\frac{3-d}{2}, \tilde{g}^*_4=[ \frac{1}{2}\pm \frac{\sqrt{\bar{c}}}{{2}}] [3-d].
\label{ring}
\end{eqnarray}
where $\bar{c}=c+1$.
Enforcing $g_{Y1}$, $g_{Y2}$ on the ring,  one can verify that Eq.\ref{ring} is infrared stable as the standard Wilson-Fisher fixed point when $\tilde{g}^*_4=[\frac{1}{{2}} +\frac{\sqrt{\bar{c}}}{2}] [3-d]$.

Meanwhile, the free-particle fixed point is defined as

\begin{eqnarray}
 \tilde{g}^*_{Y1}= \tilde{g}^*_{Y2}= \tilde{g}^*_4=0.
\end{eqnarray}
This fixed point is fully isolated but infrared unstable when $d <3$ so can't be identified as a TQCP of our interests.
In addition, there is an infrared unstable strong coupling fixed point specified as
\begin{eqnarray}
 \tilde{g}^*_{Y1}= \tilde{g}^*_{Y2}=0, \tilde{g}^*_4=3-d.
\end{eqnarray}

The ring of strong coupling fixed points shrinks towards the free-particle fixed point as one approaches the three dimension. When $d=3$, the strong coupling fixed points coalesce into the non-interacting fixed point and become marginally stable.  
On the other hand for higher dimensions or $d>3$, the only fixed point is non-interacting that is also infrared stable. However, at these fixed points, free real fermions coexist with additional emergent free or weakly interacting real scalar bosons.

Only the infrared stable fixed points in the $g_{Y1}$-$g_{Y2}$-$g_4$ space form a faithful representation of 
quantum critical points, or TQCPs. Below we will focus on the ring manifold of strong coupling fixed points. These discussions illustrate the difference between gapless TQCPs as an isolated fixed point and those as a fixed point in a smooth manifold and main applications can be found in TQCPs in low dimensional topological superconductors such as two-dimensional ones.

It is worth noting the manifold of conformal fixed points we have studied here are distinctly different from supersymmetry conformal field fixed points studied in a few previous articles\cite{Lee07,Balents98,Zerf16,Jian17,Fei16,Zhou22}. In those studies, the scalar fields that fermions are coupled to are complex ones or equivalent to complex ones rather than a single real scalar field. There is either
a $U(1)$ charge symmetry\cite{Lee07,Zerf16,Jian17,Fei16} or an emergent $U(1)$ symmetry.
In the former case, the fixed point of a $U(1)$ theory is fully isolated.
In the later case discussed in Ref.\cite{Zhou22}, the general couplings between $N_f=\frac{1}{2}$ fermions and two real scalar fields $\phi_{1,2}$ can be expressed as $g_{Y1, Y2}$. But the fixed points are again isolated with $g^2_{Y1}= \pm g^2_{Y2}=[g^*_Y]^2=\frac{1}{2}(3-d)$ reflecting an emergent $U(1)$ symmetry (when the masses of two real scalar fields vanish identically). 

The manifold here also differs from the result on supersymmetric surface states in Ref.\cite{Grover14} which requires $N_f=\frac{1}{4}$, one-half of the real fermions as in our studies. 
In all previous cases above, conformal theories are represented by standard well-isolated Wilson-Fisher fixed points but with additional emergent super-symmetries. However, in all those studies, fixe points are well isolated and do not form a smooth manifold.

\subsection{Deformations of a TQCP into conformal fields without protecting symmetry $G_p$}

We will now apply this result to illustrate what happens when a TQCP is not {\em an isolated fixed point} due to strong interactions. 
To apply to a TQCP where $M_1=M_2=0$, we again take the massless limit of $m_0=0$ so that the TQCP can be a strong coupling fixed point as the massive limit of $m_0\neq 0$ only leads to an isolated non-interacting fixed point that we have discussed in the previous sections. 

However, to have a desired time reversal symmetry of $G_p=Z^T_2$, we shall also further set $g_{Y2}=0$. In dimensions lower than three, the interacting model has two strong coupling scale invariant fixed points which represent conformal-field-theories with time reversal symmetry. They are

\begin{eqnarray}
&&\tilde{g}_{Y1}=\tilde{g}^*_Y=\pm \sqrt{\frac{3-d}{2}}, \tilde{g}_{Y2}=0, \nonumber \\
 && \tilde{g}_4=\tilde{g}^*_4=[\frac{1}{2} +\frac{\sqrt{\bar{c}}}{{2}}] [3-d].
\end{eqnarray}
The rest of the fixed points along the ring with non-vanishing coupling of $\tilde{g}_{Y2}$ all break the protecting $Z^T_2$ symmetry.

In short, in $d<3$ a strongly interacting TQCP with protecting symmetry $G_p=Z^T_2$ can now be a CFT state with strongly interacting gapless fermions and scalar bosons.
It is further smoothly connected to other strong coupling CFT states that fully break the protecting symmetry $G_p$.

Now we are ready to study what deformations do to the time reversal symmetric scale invariant fixed point.  
Applying the deformation $U_D(\lambda)$ with $D=\frac{1}{2}\int d{\bf r} \chi^T \gamma \chi$ introduced before to the Hamiltonian Eq.\ref{CFT1},  
following  Appendix C one can easily confirm explicitly that the terms involving $\beta_{1,2}$, i.e. the mass operators and Yukawa coupling operators, transform non-trivially as,

\begin{eqnarray}
\beta_1 & \rightarrow & \beta'_1=\cos\lambda\pi \beta_1+\sin\lambda\pi \beta_2, \nonumber \\
\beta_2 & \rightarrow & \beta'_2=- \sin\lambda\pi \beta_1+\cos\lambda\pi \beta_2. 
\end{eqnarray}
while the rest of the terms simply remain invariant.

The deformations of $U_D(\lambda)$ thus lead to the following transformations of the CFT fixed point for the TQCP with the protecting symmetry $G_p$,

\begin{eqnarray}
&& {H}_{eff} (0; g_{Y1}, g_{Y2}; g_4)=U^\dagger_D(\lambda) H_{eff}(0; g^*_{Y}, 0; g^*_4)U_D(\lambda) \nonumber \\
&& g_{Y1}=\cos\lambda \pi g^*_{Y}, g_{Y2}=-\sin\lambda\pi g^*_{Y}, g_4=g^*_4.
\end{eqnarray}
As in gapped phases, the deformations break the protection symmetry $G_p=Z^T_2$ and lead to CFTs along the ring without the protecting symmetry $G_p$. For instance, at $\lambda \pi=0,\pi$, the CFTs are time reversal invariant as $g_{Y2}=0$ while 
all other gapless CFTs break the time reversal symmetry as $g_{Y2} \neq 0$.
And the Yukawa coupling follows a simple circular rule under the deformations

\begin{eqnarray}
g^2_{Y1}+g^2_{Y2}=[g^*_{Y}]^2, g_4=g_4^*.
\end{eqnarray}
Deformations $U_D(\lambda)$ transform the strong coupling TQCP represented by a CFT with $g_{Y1}=g^*_Y$, $g_{Y2}=0$ and $g_4=g^*_4$ into other strong coupling fixed points along the ring with symmetries lower than $G_p$.

So in this case, the fixed point is not isolated and the actual strong coupling CFT fixed points form a ring which can be precisely traced out by the deformations $U_D(\lambda)$. 
Gapless CFT states surrounded by gapped phases are not unique.
The smooth deformations in this case do not lead to an emergent continuous symmetry in the gapless TQCP, which is in stark contrast to the weakly interacting fixed point where there is a unique free-fermion gapless state related to a fully isolated fixed point
in the space of interacting Hamiltonians.

 As mentioned before, in addition to deformations of a gapped topological state into gapped state without protecting symmetry $G_p$, $U_D(\lambda)$ also deforms the strong coupling fixed point of TQCP with symmetry $G_p$ into other fixed points with lower symmetries. In fact, the manifold can exactly be traced out by the deformations $U_D(\lambda)$ and represents a set of conformal-field-theory fixed points without protecting symmetries except the one connected to TQCPs.

\subsection{Duality and Emergent $Z_2$ Symmetry}

Although the deformation group $U_D(\lambda)$, $\lambda\in [0,2]$ generally breaks the protecting symmetry $G_p$ and can transform a TQCP into conformal fields in a lower-symmetry manifold, there is a subgroup of $U_D(\lambda)$ that is
invariant under $G_p$.This property of deformations is essential because for gapped states there shall be at least two states along the deformation path that have the protecting symmetry $G_p$ (See section III and Fig.\ref{TQCPEMS}).
This property ensures that a path traced out by continuous deformations can connect two topologically distinct gapped states symmetry with protecting symmetry $G_p$. 
In our case, we define it as $Z_2$ subgroup defined by \{ $U_D(\lambda=0)$, $U_D(\lambda=1)\}$. 
For gapped states $|\Psi_{T0, T1} \rangle$, $|\Psi_{T1} \rangle =U_D(\lambda=1)|\Psi_{T0}\rangle$.

 In terms of weakly interacting EFTs in Eq.\ref{Eff} in Section VI, one can verify that 

\begin{eqnarray}
&&H_{eff}(\pm M_1, M_2=0) \rightarrow H_{eff}(\mp M_1, M_2=0) \nonumber \\ 
&&=U^{-1}_D(\lambda=1)H_{eff}(\pm M_1, M_2=0)  U_D(\lambda=1);\nonumber \\
&& U^{-1}_{G_p}  U_D(\lambda=1) U_{G_p}  = U_D(\lambda=1).
\end{eqnarray}
which indicates a duality between two EFTs on two sides of a TQCP and for $U_D(2)=U_D^2(1)$, $U_D^{-1}(2) H_{eff} U_D(2)=H_{eff}$. That is,
as $U_D(\lambda=1)$ is invariant under the protecting symmetry group $U_{G_p}$, the Hamiltonian of the EFT in one topological phase is {\em dual} to the one in the other phase under the same protecting symmetry $G_p$.

Now if the TQCP is an isolated fixed point one, when $M_1=M_2=0$, 

\begin{eqnarray}
&&H_{eff}(M_1=0, M_2=0)  \rightarrow  H_{eff}(M_1=0, M_2=0) \nonumber \\
&&= U_D^{-1}(\lambda=1) H_{eff} (M_1=0, M_2=0)  U_D(\lambda=1), 
\end{eqnarray}
and 
\begin{eqnarray}
 U_D^{-1}(\lambda=1) \chi  U_D(\lambda=1) &=& i\gamma \chi.
\end{eqnarray}
The gapless TQCP, if is an isolated fixed point, needs to be self-dual under $U_D(\lambda=1)$. The self-duality in this case is sufficient to leave the gapless state invariant under any of $U_D(\lambda)$, $\lambda \in [0,2]$ leading to an emergent $U(1)$ symmetry. i.e.
 
\begin{eqnarray}
&& H_{eff}(M_1=0, M_2=0) \rightarrow H_{eff}(M_1=0, M_2=0) \nonumber \\
&&=U^{-1}_D(\lambda) H_{eff} (M_1=0, M_2=0)  U_D(\lambda),
\end{eqnarray}
and 

\begin{eqnarray}
 U^{-1}_D(\lambda)  \chi U_D(\lambda) &=& \exp(i\frac{1}{2}\lambda\pi\gamma) \chi, \lambda \in [0,2].
\end{eqnarray}
Consequently, the self-duality of an isolated fixed point therefore leads to a continuous emergent symmetry.

Let us now turn to the strong coupling limit and CFT fixed points.
A TQCP can be specified as one of the CFTs with $g_{Y1}=\pm g^*_Y$, $g_{Y2}=0$ in Eq.\ref{CFT1}. These are the only fixed points with the protected symmetry $G_p$. The action of $U(\lambda=1)$ transforms one into the other and vice versa.

\begin{eqnarray}
&&H_{eff}(0; \pm g^*_{Y}, 0; g^*_4)\rightarrow \nonumber \\
&&{H}_{eff}(0; \mp g^*_{Y}, 0; g^*_4)=
U_D^{-1}(1) H_{eff}(0; \pm g^*_{Y}, 0; g^*_4)U_D(1).\nonumber \\
\label{dual}
\end{eqnarray}
Therefore, the two $G_p$-symmetric CFTs related by $U_D(\lambda=1)$ are dual to each other under the deformation transformation $U_D(\lambda=1)$. And $U^2_D(\lambda=1)=U_D(\lambda=2)$, which is an identity when acting on the CFTs.

This dual relation leads to the following $Z_2$ parity symmetry at the TQCP as the CFTs in Eq.\ref{CFT1} indeed remain symmetric under a $Z_2$ parity symmetry transformation. That is,

\begin{eqnarray}
&& \chi ({\bf r}) \rightarrow \chi'({\bf r})=\exp(i\frac{\pi}{2} \gamma) \chi({\bf r}),
\phi({\bf r}) \rightarrow -\phi({\bf r}). \nonumber \\
&& H_{eff}(0;g^*_Y, 0; g^*_4)  \rightarrow H'_{eff}=H_{eff} (0;g^*_Y, 0;g^*_4).
\end{eqnarray}
This duality symmetry is the discrete emergent symmetry surviving a smooth manifold of many fixed points. So in the strong coupling limit, a TQCP has an emergent $Z_2$ symmetry rather than a continuous $U(1)$ symmetry in the weakly interacting limit.
 The symmetry group of the TQCP is $H=Z^T_2 \otimes Z_2$ with an emergent symmetry related to the duality in Eq.\ref{dual}.

\section{Conclusions}

To summarize, we have established an intimate connection between emergent symmetries in gapless states or TQCPs with protecting symmetry $G_p$, and continuous deformations $U_D(\lambda)$
in surrounding gapped phases with no protection symmetries. We show that if the TQCPs are represented by a fully isolated scale invariant fixed point, smooth deformations shall lead to an emergent continuous symmetry $U_{EM}$
with $U_{EM}$ precisely given by $U_D(\lambda)$.

However if TQCPs are one of CFT fixed points located in a smooth manifold of other fixed points that do not have protecting symmetry $G_p$, continuous deformations don't leave TQCPs invariant. Instead, a subgroup of the deformation
group $U_D(\lambda)$ that is symmetric under the group of protecting symmetry $G_p$ transforms a fixed point of TQCP to into its dual.
This results in a duality symmetry between CFT fixed points with protecting symmetries. The TQCP in the strong coupling limit has an emergent duality symmetry or a $Z_2$-parity symmetry although gapped states around the TQCP are still continuously deformable.

We also explicitly elevate the $d$-dimension lattice model of $N_f$-fermions with protecting symmetry $G_p$ to  $(d+1)$-dimension lattice models of $2N_f$ fermions
with higher protecting symmetry $H=G_p\ltimes U_{EM}$. Using the method of elevation of symmetry, fermion number along with the dimension, we illustrate that the $d$-dimension TQCPs between topological states with protecting symmetry $G_p$ can be properly represented by a surface of $(d+1)$-dimension lattices with higher protecting symmetries $H$
and two surfaces correspond to two copies of TQCPs under parity transformations. In the case of symmetry group $G_p=Z^T_2$, our analysis shows a $3D$ TQCP in topological states of DIII class without charge $U(1)$ symmetry can be properly represented by a surface of a $4D$ lattice model that is dual to $4D$ {\em strong topological insulators} with charge $U(1)$ symmetry.

Although in this article, we mainly focus on $3D$ TQCPs in topological superconductors, the phenomenologies of deformations and their relations to emergent symmetries are very generic.
While details of emergent symmetries remain to be understood in many other phenomena, we also believe the approaches outline here can be useful for identifying symmetries or deformations of gapped states around a TQCP. 
For instance, in a celebrated $1D$ model where AKLT-like states can make a transition to trivial states without edge degeneracies\cite{Pollman12}, a {\em gapless} topological quantum critical point has been explicitly constructed and further displays an emergent symmetry of $SU(2)$ so that the state is further invariant under $SU(2) \otimes SU(2)$ rather than $SU(2)$ itself as in the gapped phases. The deformations induced by the emergent $SU(2)$ symmetry group potentially can be a way to understand how gapped states in two phases around  a critical point can be smoothly deformed into each other.

The studies here rely on identifications of smooth deformations of gapped or CFT fixed points that lead to emergent symmetries, either continuous or discrete. All the deformations we have considered are generated by {\em local unitary transformations} defined by $D$. The question of how to generalize these discussions to {\em deformations} of non-local transformation remains to be investigated.

In earlier studies of $1D$ gapless states, quantum doubling taking into account dual-theory symmetries further play an extremely important role, 
contributing to more emergent symmetries even in conventional order-disordered transitions\cite{Ji20}. In high dimensions, dual-theory symmetries are usually characterized by categorical symmetries or high-form extended symmetries. In high spatial dimensions we have focused on, how those emergent high-form symmetries look like in simple deformation pictures remain to be fully understood and can be an exciting direction to pursue in the future.  It is possible that in those cases one needs to take into account deformations with {\em non-local unitary transformations} beyond what we have considered here. These generalizations, if possible, can be very inspiring for better understanding of general categorical symmetries.
The other open questions are the boundary dynamics of gapless states discussed here, whether there can be non-trivial surfaces even when the bulk becomes critical. From the point of view of anomalies, it won't be impossible to have fascinating boundaries in the strong coupling CFT limit. However, this remains to be seen in the future.

As mentioned in the introduction, it is interesting to further explore the physical consequences of emergent symmetries in gapless states from a point of view of observations.
Can one actually probe such emergence in direct measurements? In general, conformal field theories (CFT) crucially depend on underlying symmetries. How the emergent symmetries shall be part of the construction of CFTs remains to be debated.
An earlier analysis seems to suggest that emergent symmetries can be quite restricted in the infrared limit and it can quickly evolve into lower symmetry or no symmetries at all once moving up in the energy scale. Its relevance in the strong coupling limit can be further debated.

 The other fascinating question is what is the use of emergent symmetries? Can they be applied to trap or manipulate emergent topological particles or even topological qubits? In the class of TQCPs studied here in the weakly interacting limit,  the dynamics can be mapped into a single copy of Weyl fermions which are usually forbidden in the presence of charge $U(1)$ symmetry. Consequently, the emergent symmetries are not gauge-able because of the well known t'Hooft anomalies.
 We further notice that there are no standard Adler-Bell-Jackiw type mixed anomalies because in the effective field theory here the real fermions don't interact with the charge $U(1)$  gauge fields. Absence of emergent gauge fields at generic TQCPs
in charge-$U(1)$ symmetry-breaking superconductors turns out to be a very unique feature, that differs from that at de-confined quantum critical points emphasized previously\cite{Senthil04}. 
However, we don't expect that this remains to be a generic feature at TQCPs between topological phases with charge $U(1)$ symmetries. We plan to further research this subject in the near future.

The author wants to thank Julio Parra-Martinez for a discussion on marginal operators around conformal manifolds and Fan Yang for discussions on applications to states in the CI class. This project is in part supported by NSERC (Canada) Discovery Grant under the contract No RGPIN-2020-07070.

\appendix

\section{Continuous deformations forbbidened around a QCP in the Landau Paradigm}

Let us consider an order-disordered quantum phase transition in high dimensions with symmetry $G$.
For complex fields $\phi({\bf r})$ and conjugate fields $\pi({\bf r})$, we introduce the following general model

\begin{eqnarray}
H &=& H_0 +H_M +H_I \nonumber \\
H_0&=& \int d{\bf r} [ \pi^* ({\bf r}) \pi({\bf r})+\nabla \phi^* ({\bf r}) \cdot \nabla \phi({\bf r}) ], \nonumber \\ 
H_M &=& \int d{\bf r} [ m_0 \phi^*({\bf r}) \phi({\bf r}) + \frac{m_1}{2} ( \phi^*({\bf r}) \phi^*({\bf r})+ \phi({\bf r}) \phi({\bf r}) )\nonumber\\
&&+i \frac{m_2}{2}  (\phi^*({\bf r}) \phi^*({\bf r})- \phi({\bf r}) \phi({\bf r}) )
\end{eqnarray}
where we have muted interaction terms in $H_I$. And $[\phi({\bf r}), \pi^*({\bf r}')]=i \delta({\bf r} -{\bf r}')$. 

The  mass operator of $m_0$ is $U(1)$ symmetric and is the most symmetric one among all the three mass operators above.
The operator of $m_1$ breaks the $U(1)$ symmetry group but remains invariant under both a $Z_2$ subgroup of $\phi \rightarrow -\phi$ and time reversal transformation
$\phi \rightarrow \phi^*$. The operator of $m_2$ breaks both the time reversal symmetry of $Z_2^T$  and $U(1)$ symmetry but is still invariant under the $Z_2$ reflection.

Therefore when both $m_1$ and $m_2$ are zero, the symmetry group is $G=U(1) \rtimes Z^T_2$ while when $m_1$ is finite but $m_2=0$, the model is symmetric under symmetry group $G=Z_2 \otimes Z^T_2$.
When $m_2$ also becomes finite, the model has the lowest symmetry with $G=Z_2$.
For our purpose, it is convenient to re-express it  in the following representation,

\begin{eqnarray}
H &=& H_0 +H_M +H_I \nonumber \\
H_0&=& \int d{\bf r} [ \pi^T({\bf r}) \pi({\bf r})+\nabla \phi^T({\bf r}) \nabla \phi({\bf r}) ], \nonumber \\ 
H_M &=& \int d{\bf r} [ m_0 \phi^T({\bf r}) \phi({\bf r}) + m_1 \phi^T({\bf r}) \tau_z \phi({\bf r}) \nonumber\\
&&+m_2 \phi^T({\bf r}) \tau_x \phi({\bf r}) ]
\end{eqnarray}
where $H_I$ are again symmetric interactions.
$\pi$ is again a conjugate field of $\pi^T=(\pi_1,\pi_2)$ and satisfies the standard commutation relation $[\phi({\bf r}), \pi^T({\bf r}')]=i\delta({\bf r}-{\bf r}') \mathcal{I}$.
Finally, $\phi^T=(\phi_1,\phi_2)$ is a two-component real scalar field which can be thought as real and imaginary parts of the complex field. 

\begin{eqnarray}
\phi=\phi_1 +i\phi_2, \phi^*=\phi_1-i\phi_2
\end{eqnarray}
$m_{0,1,2}$ carry the units of the square of masses but can be either positive or negative.

In the limit of $m_1=m_2=0$, the model is symmetric under the group, $G=U(1) \rtimes Z^T_2$. 
A finite mass term of $m_1$ breaks the $U(1)$ symmetry and lowers the symmetry to $G_S=Z_2 \otimes Z^T_2$.
Here $Z_2$ is the reflection symmetry and $Z^T_2$ is the time reversal transformation in the symmetry group $G$.  When $m_2=0$, these symmetries can be defined as

\begin{eqnarray}
\mbox{$Z_2$:} \phi \rightarrow \tilde{\phi} &=&- \phi, H\rightarrow \tilde{H}=H; \nonumber\\
\mbox{$Z^T_2$:} \phi \rightarrow \tilde{\phi} &=& K \tau_z \phi, H\rightarrow \tilde{H}=H.
\end{eqnarray}
$K$, a complex conjugate, can appear in $Z^T_2$ transformation but in the real scalar field representation, $K=1$ as all matrix operators in the Hamiltonian have to be real hermitian.
When $m_2$ becomes nonzero, $Z^T_2$ subgroup in the symmetry group $G_S=Z_2 \otimes Z_2^T$ becomes further broken\cite{Z_2}. 

In the $m_0-m_1$ plane when $m_2=0$ or the $m_0-m_2$ plane when $m_1=0$,
moving away from the axis $m_0$ where $m_1$ or $m_2$ becomes nonzero lowers the symmetry $G$ to $G_S=Z_2\otimes Z^T_2 \subset G$ or $G_S =Z_2 \subset G$. 
The quantum critical point (QCP) of $m_0=0$ in the model with $G=U(1) \rtimes Z^T_2$ becomes an intersection of critical lines $m_0 =\pm m_1$ in the $m_0-m_1$ plane when $m_2=0$ or critical lines of $m_0 =\pm m_2$ in the 
$m_0-m_2$ plane where $m_1=0$. And there can be no smooth deformations connecting order and disordered phases along the horizontal direction of $m_0$ where $m_1=m_2=0$ and $G=U(1) \rtimes Z^T_2$ (See Fig.\ref{LandauPD}).

Let us now restrict to the plane of $m_0-m_1$ where $m_2$ is set to be zero.
The $Z^T_2$ transitions are given by a straight line of $m_1 =m_0$ and $Z_2$ transition line is $m_1=-m_0$(See Fig.\ref{LandauPD}).
So for a generic theory where $m_0 \neq 0$ and positive, there shall be two order-disordered transitions driven by $m_1$. At $m_1=m_{c1}=m_0$, there is a transition between symmetric phase and $Z^T_2$ time reversal symmetry breaking phase.
At $m_1=m_{c2}=-m_0$, there is a transition between symmetric phase and $Z_2$ reflection symmetry breaking phase.  And $m_{c1} > m_1 > m_{c2}$, the phase is disordered with symmetry of $G_S=Z_2 \otimes Z^T_2$ or $H=G_S=Z_2 \otimes Z^T_2$.
Beyond the critical points of $m_{c1,c2}$, ordered states are invariant under subgroups $H_1=Z_2*Z_2^T$, $H_2=Z_2^T$ respectively. Here $Z_2*Z_2^T$ refers to a symmetry under the combined of two transformations.
And $H_{1,2} \subset G_S$ as in the standard paradigm. At the critical points, as usual, $H=G_S$. 
One can extend easily extend this result to the $m_0-m_2$ plane when $m_1=0$.

In the $m_1-m_2$ plane but with non-zero $m_0 (>0)$,
the extension of $m_{c1,c2}$ turns into a close circle of radius $m_1^2+m_2^2=m_0^2$ which is one of the general scenarios discussed in the main text (See Fig.\ref{LandauPD}).
Within the circle, states are fully gapped and disordered. Outside are smoothly connected symmetry broken states. However, there can be no smooth deformations encircling around either of $m_{c1,c2}$ points as both are on the circular line of critical points
in the two dimensional plane. In the order-disorder transition paradigm, this hinders smooth deformations that had been found around a TQCP.

It is for this reason, generically we don't have continuous emergent symmetries around these QCPs along the $m_1$- or $m_2$-axis with symmetry $G_S$. At gapless QCPs, the symmetry group $H$ is exactly $G_S$, not higher that $G_S$.

 \section{Deformations of a gapped state around a gapless phase  with protecting symmetry $G_p$}

The gapped nodal states when $B^{eff}_x=B_0$ can be expressed as

\begin{eqnarray}
|\Psi_0 \rangle = \Pi_{\{{\bf k},-{\bf k}\}} f^\dagger_{{\bf k}, +} f^\dagger_{- {\bf k}, +} 
 f^\dagger_{{\bf k},-} f^\dagger_{- {\bf k}, -} |0\rangle
\end{eqnarray}
where $f^\dagger_{{\bf k, +}}$ and $f^\dagger_{{\bf k, +}}$ are creation of fermionic modes with negative frequencies near $L, R$ nodal points respectively (See below).

Mode operators $f^\dagger_{{\bf k}, +}$, $f^\dagger_{{\bf k}, -}$ are related to the fermion creation operators ${\Psi^\dagger}^T=(\Psi^\dagger_S,\Psi^\dagger_A)$ via the standard BdG form.

\begin{eqnarray}
f^\dagger_{{\bf k}, +} &=& {u_+} \Psi^\dagger_{{\bf k}, +} +  v_+({\bf k}) \Psi_{-{\bf k}, +},  \nonumber \\ 
f^\dagger_{{\bf k}, -} &= &{u_-} \Psi^\dagger_{{\bf k}, -} +  v_-({\bf k})  \Psi_{-{\bf k}, -}.
\end{eqnarray}

In this part of discussions, we have defined that 
$\Psi^\dagger_{S,A}({\bf r}_0)$ are local eigen operators of $Q_z=\int d{\bf r} \Psi^\dagger\sigma_z\Psi$ with eigenvalues $\pm$ respectively. Furthermore, the "$\pm$" operators are defined as

\begin{eqnarray}
\Psi^\dagger_{{\bf k}, +} &=&  [Q_{\bf n}, \Psi^\dagger_{{\bf k}, +}], \Psi_{-{\bf k}, +}  =- [Q_{\bf n}, \Psi_{-{\bf k}, +}];
 \nonumber \\ 
 \Psi^\dagger_{{\bf k}, -}&=&- [Q_{\bf n}, \Psi^\dagger_{{\bf k}, -}],
 \Psi_{-{\bf k}, -} =  [Q_{\bf n}, \Psi_{-{\bf k}, +}].
\end{eqnarray}
That is,
$\Psi^\dagger_{{\bf k},\pm}$ are the eigen operators of $Q_{\bf n}$ with eigen values $\pm$ respectively. And

\begin{eqnarray}
&&Q_{\bf n}=
\int d{\bf r} \Psi^\dagger({\bf r}) {\bf \sigma} \cdot {\bf n} \Psi({\bf r}).
\label{sigma_z}
\end{eqnarray}
The unit vector ${\bf n}$ is defined along the direction ${\bf B}^{eff}$,

\begin{eqnarray}
{\bf n}=\frac{{\bf B}^{eff}}{| {\bf B}^{eff} |}, {\bf B}^{eff}=(B_0, ck_y, 0).
\end{eqnarray}

The deformation operator is

\begin{eqnarray}
D=\frac{1}{2}\int d{\bf r} \Psi^\dagger({\bf r}) \sigma_y \Psi ({\bf r}), [D,\Psi^\dagger({\bf r}_0) ]= \Psi^\dagger({\bf r}_0) \sigma_y \nonumber \\
\label{sigma_y} 
\end{eqnarray}

Following Eq.\ref{sigma_z}, and

\begin{eqnarray}
U_D^\dagger (\lambda) \Psi^\dagger({\bf r}_0) U_D(\lambda)&=& \Psi^\dagger({\bf r}_0)\exp(i\frac{1}{2}\lambda \pi \sigma_y),  \nonumber \\
U_D^\dagger (\lambda) \Psi({\bf r}_0) U_D(\lambda)&=&\exp(-i\frac{1}{2}\lambda \pi \sigma_y) \Psi({\bf r}_0),
\end{eqnarray}
one can show that $U (\lambda) |\Psi_0 \rangle$ with $ \lambda \in [0,2]$ is equivalent to the ground state of ${\bf B}^{eff}$ rotated around the $y$-axis by an angle of $\lambda \pi$ starting at $B^{eff}_x=B_0$ and $B^{eff}_z=0$. 

Indeed, under $U_D(\lambda)$,
\begin{eqnarray}
Q_{\bf n} &\rightarrow& Q_{\tilde{\bf n}}, \tilde{{\bf n}}_y = {\bf n}_y,\nonumber\\
\tilde{{\bf n}}_x &= &{\bf n}_x \cos\lambda \pi+{\bf n}_z \sin\lambda \pi, \nonumber \\
\tilde{{\bf n}_z} &= &{\bf n}_z \cos\lambda\pi -{\bf n}_x \sin\lambda \pi. \nonumber \\
\end{eqnarray}

In the real fermion representation of $\chi^\dagger=\chi^T$, the effective field theory for nodal phases can be written as 

\begin{eqnarray}
H_{eff} &=& H_0+\sum M_m \chi^T \beta_m\chi  \nonumber \\
H_0 &=&
 \frac{1}{2} \int d{\bf r} [ \chi^T({\bf r}){\bf  \alpha} \cdot {i\nabla} \chi ({\bf r})
\nonumber \\
&+& g_1 \chi^T \beta_1 \chi \chi^T  \beta_1 \chi +g_2 \chi^T  \beta_2 \chi  \chi^T \beta_2 \chi +... ] \nonumber \\
\label{Eff1}
\end{eqnarray}
where ${\bf \alpha}=(\alpha_x,\alpha_y,\alpha_z)$. $\alpha_{i}$, $i=x,y,z$ and $\beta_{1,2}$ are mutually anti-commuting Hermitian matrices having the following explicit forms

\begin{eqnarray}
\alpha_x &=& \tau_x\otimes I, \alpha_z=\tau_z\otimes 1,\alpha_y= {c} \tau_y\otimes \sigma_y \nonumber \\
\beta_1&=& \tau_y \otimes \sigma_z, \beta_2=\tau_y \otimes \sigma_x.
\end{eqnarray}

\section{Deformation operator in terms of Real-fermion fields of EFT}

Consider $2n$ component real fermion fields $\chi({\bf r})$.
For a general deformation operation  of the form 

\begin{eqnarray}
D=\frac{1}{2} \int d{\bf r} \chi^T \mathcal{D}\chi, \mathcal{D}^T=-\mathcal{D}=\mathcal{D}^*
\end{eqnarray}
Under $U_D(\lambda)=\exp(i\frac{1}{2} \lambda \pi D), \lambda \in [0,2]$,

\begin{eqnarray}
U_D^\dagger(\lambda) \chi({\bf r}_0) U_D(\lambda) &=& \exp(-i\frac{1}{2}\lambda \pi \mathcal{D})\chi({\bf r}_0), \nonumber \\
U_D^\dagger(\lambda) \chi({\bf k}) U_D(\lambda) &=&\exp(-i\frac{1}{2}\lambda \pi \mathcal{D})\chi({\bf k}).
\end{eqnarray}
$U_D$ can be any $U(1)$ subgroup in a Spin(2n) group which all together has $n(2n-1)$ generators with group algebras isomorphic to $SO(2n)$. This property has been applied to track the deformations of gapped states or EFTs.

\section{$d$-dimension surfaces of $(d+1)$-dimension Bulk with higher lattice symmetries}

This section is to explicitly construct a $(d+1)$-dimension lattice model, the surface of which can realize an EFT for $d$-dimensional TQCPs with $U_{EM}=U(1)$ emergent symmetry and protecting symmetry $G_p=Z^T_2$.
The $(d+1)$-dimension lattice bulk turns out to be, effectively, a $4D$ {\em strong topological insulator} with charge $U(1)$ symmetry. An equivalence is established below between a $3D$ TQCP with $N_f=\frac{1}{2}$ gapless fermions in a DIII class 
topological states without charge $U(1)$ symmetry and a $3D$ surface of a $4D$ strong topological insulator with a charge $U(1)$ symmetry of $N_f=1$ fermions.

\subsection{What happens to deformation operators in a $(d+1)$D gapped bulk?}

Now we will construct the $(d+1)$-dimension model with lattice symmetry $H=G_p \ltimes U(1)$, starting with a $d$-dimension model with lattice symmetry equal to $G_p=Z^T_2$ but without $U(1)$ symmetry. 
The main idea is to enlarge the symmetry group by increasing fermion numbers, in this case simply by doubling $N_f=\frac{1}{2}$ to $N_f=1$. For $d=3$,

\begin{eqnarray}
H^{(d+1)}&=&H^{(d+1)}_0 +H^{(d+1)}_M,\nonumber \\
H^{(d+1)}_0&=&\sum_{{\bf n}, {\bf a}} \chi^T_{\bf n} i {\bf \alpha}\cdot {\bf a} \chi_{{\bf n} +{\bf a}} +g_0\sum_{\bf n} \chi^T_{\bf n} \beta_0\chi_{\bf n}  \chi^T_{\bf n} \beta_0\chi_{\bf n}, \nonumber \\
H^{(d+1)}_M&=&2 M_0 \sum_{{\bf n},{\bf e}_\eta} \chi^T_{\bf n} i \beta_0\otimes \Sigma_y \chi_{{\bf n}+{\bf e}_\eta}  \nonumber \\
&-& M_{1A}\sum_{{\bf n},{\bf b}} \chi^T_{\bf n} \beta_0\otimes \Sigma_x \chi_{{\bf n}+{\bf b}} +
 M_{1B} \sum_{{\bf n}} \chi^T_{\bf n} \beta_0\otimes \Sigma_x \chi_{\bf n}. \nonumber \\
\label{4D}
\end{eqnarray}
Here $H^{(d+1)}_0$ is identical to $H_0$ defined in Eq.\ref{3Dlattice} but with summations over $4D$ lattice sites, ${\bf n} =(n_x, n_y, n_z, n_\eta)$, $n_i$, $i=x,y,z,\eta$ are integers.
Unit vectors ${\bf a}=\pm {\bf e}_i$ are along the positive and negative $i=x,y,z$-directions respectively and unit vectors ${\bf b}$ are along the positive and negative of $x,y,z,\eta$-directions.

$\Sigma_{x,y,z}$ are Pauli matrices for the quantum doubling.
The first term in $H^{(d+1)}_M$ is proportional to $M_0$; the sum is along the positive $\eta$-th or 4th direction only and ${\bf e}_\eta$ is the unit vector pointing to a neighbouring site along that direction.  
$M_{1A}$, $M_{1B}$ are the mass terms playing an important role in discussions of $3D$ surfaces which we will turn to shortly. 
In addition, 

\begin{eqnarray}
[\beta_0 \otimes \Sigma_y]^T=\beta_0 \otimes \Sigma_y,
[\beta_0 \otimes \Sigma_x]^T=-\beta_0 \otimes \Sigma_x.
\end{eqnarray}

The $Z^T_2$ symmetry is evident as $\beta_0$ is a time-reversal symmetric operator and $i\Sigma_y, \Sigma_x$ are real matrices,

\begin{eqnarray}
\mathcal{T}^{-1} H^{(d+1)}\mathcal{T}=H^{(d+1)}.
\end{eqnarray}

Consider $U_{d+1}(\lambda)$ generated by a $(d+1)$-dimension deformation operator,

\begin{eqnarray}
D_{d+1}=\frac{1}{2}\int d{\bf r} \chi^T  q_1 \otimes \Sigma_z \chi , U_{d+1}=\exp(i \frac{1}{2}\lambda\pi D_{d+1}). \nonumber \\
\end{eqnarray}
This deformation operator $D_{d+1}$ differs from the $d$-dimension one, $D$, introduced in Section VI by $\Sigma_z$ operator.

The $(d+1)$-dimension lattice model is now invariant, i.e. under $U_{d+1}(\lambda)$,

\begin{eqnarray}
{\chi} &\rightarrow& \tilde{\chi} = U_{d+1}^{-1} (\lambda) \chi U_{d+1}(\lambda), \nonumber \\ 
H^{(d+1)} & \rightarrow& \tilde{H}^{(d+1)}=U_{d+1}^{-1} H^{(d+1)} U_{d+1}=H^{(d+1)}
\end{eqnarray}
so that indeed the lattice model has a higher symmetry of $H=Z_2^T \ltimes U(1)$ instead of $Z^T_2$.

Below we will show that on surfaces, $D_{d+1}$ and $U_{d+1}$ deformations fall back to desired $3D$ deformation operator $D$ on two opposite surfaces. 
The EFTs on two surfaces are precisely related to two TQCPs with emergent symmetries and are simply connected by a parity transformation.
Two deformations are smoothly connected to 4D bulk deformations.

\subsection{$d$-dimension surfaces of $(d+1)$-dimension Gapped bulk as EFT of TQCPs} 

Now we will find the $3D$ surface EFTs for the $4D$ bulk with lattice symmetry given by $H=Z^T_2 \ltimes U(1)$ and identify them as TQCPs in $3D$ with protecting symmetry $G_p=Z_2^T$ and emergent symmetry $U_{EM}=U(1)$ or TQCPs in states of DIII class.

If we set $M_{1B}=0^+$ and focus on the momentum near $k_\eta=0$, we can obtain the typical domain-wall fermions in a $3D$ surface. 
(To prevent the bulk from developing gapless nodal points, we enforce that $M_{1B}$ is a small positive value.)
The surface projection operator is 

\begin{eqnarray}
P_s(\pm)=\frac{1\pm\Sigma_z}{2}, 
\end{eqnarray}
 $P_s^2(\pm)=1$ precisely projecting out the desired gapless $N_f=\frac{1}{2}$ fermions on the surface.

However, on the surface, there will be $2^{3-1}$ gapless $N_f=\frac{1}{2}$ Dirac fermions around $(k_x,k_y,k_z)=(\xi_1,\xi_2,\xi_3)\pi$ points in the momentum space, $\xi_i=0,1$, $i=1,2,3$.
They are all invariant under the $d$-dimension deformations $U_D(\lambda)$ as on the surface $\chi_s=P_s(\pm)  \chi$, $\Sigma_z P_s(\pm)=\pm P_s(\pm)$. The surface states are eigen states of $\Sigma_z$.

Especially, if the surface is projected out by operator $P_s(+)$, the $(d+1)$-dimension deformation operator $D_{d+1}$ becomes exactly equal to $D$ and falls back to be the 
$d$-dimension deformation operator discussed in the previous section. In such surfaces, $U_{d+1}(\lambda)=U_D(\lambda)$, becomes fully equivalent to the $3D$ deformations on the surface.  So in this case, there will still be $2^{3-1}$ copies of $N_f=\frac{1}{2}$ gapless fermions, leading to $N_f=2$ effective Dirac fermion fields, half of the degrees in bulk $3D$ lattices with the same lattice symmetry where there can be $2^3$ copies of $N_f=\frac{1}{2}$ gapless fermions.  And just like in the 3D lattices, in this construction, half of these fermions are left-handed and others are right handed in a $3D$ surface.

The large number of copies of $N_f=1/2$ gapless fermions can be further reduced by working with the model of a finite $M_{1B}$ without lowering the symmetry group of $H=Z^T_2 \ltimes U(1)$.
We can work with a mass so that $8 M_{1A} > M_{1B} > 4M_{1A}$. This leads to a negative mass at momentum $(0,0,0,0)$ only  but positive ones for the rest of Dirac cones at $(\xi_1,\xi_2,\xi_3,\xi_4)\pi$, $\xi_i=\pm 1$, $i=x,y,z,\eta$. 
The gapless surface state of such a 4D gapped bulk with $H=Z^T_2 \ltimes U(1)$ therefore will have exactly $N_f=\frac{1}{2}$ fermions and naturally is symmetric under $U_D(\lambda)$ or the $U(1)$ group, apart from being symmetric under $Z_2^T$.

The effective field theory of the $4D$ bulk for our studies of $3D$ surfaces has a simple form of

\begin{eqnarray}
H^{(d+1)}_{eff} &=&
 \frac{1}{2} \int d{\bf r} [ \chi^T({\bf r})[{\bf  \alpha} \cdot {i\nabla} +M_0 \beta_0\otimes \Sigma_y \cdot i\partial_\eta] \chi ({\bf r})
\nonumber \\
&+&  M_{1} \int d{\bf r} \chi^T({\bf r}) \beta_0 \otimes \Sigma_x  \chi({\bf r})+... \nonumber \\
\end{eqnarray}
where $M_1=M_{1A}-8M_{1B}$ is a negative mass.

As mentioned before, for the surface with projection operator $P_s(+)$, the deformation operator becomes $D_{d+1}=D$, the $d$-dimension operator.  
The opposite surface in 4D has a projection operator $P_s(-)=\frac{1-\Sigma_z}{2}$, $\Sigma_z P_s(-) =-P_s(-)$. The $4D$ deformation operator $D_{d+1}=-D$, falls back to be the $3D$ one with an additional minus sign. 
$U_{d+1}(\lambda) =U_D(-\lambda)$, instead being equal to $U_D(\lambda)$ itself. This is equivalent to make a transformation from $q_1 \rightarrow - q_1$ with $q_1=i\alpha_x\alpha_y\alpha_z$. 
Therefore, the $U(1)$ symmetries on two surfaces are given by $D$, $U_D(\lambda)$ and $-D$, $U_D(-\lambda)$ respectively, a characteristic of anomalies.

On the physic side, the opposite surface represents the EFT for another gapless quaternion superconductor with $\Delta_0 \rightarrow -\Delta_0$ and $n_s=0$ or a TQCP so that it is symmetric under $Z^T_2$. The two $3D$ TQCPs represented on two opposite surfaces in $4D$ are simply connected by a reflection of ${\bf k} \rightarrow -{\bf k}$ in their $3D$ spaces. It is further tempting to point out the $4D$ gapped bulk is also a topological state but with higher protecting symmetry of $Z^T_2 \ltimes U(1)$. In fact, one can identify it as a dual of a $4D$ topological insulator with charge $U(1)$ symmetry but in a real fermion representation.

Note that this construction of $3D$ real fermions does not lead to single chirality of fermions, unlike in standard quantum-filed-theory approaches to single chirality complex Dirac fermions. The reason is that at TQCPs we are interested in have both left- and right-handed real fermions. What we have achieved is to produce a surface with only one single cone of gapless real Dirac fermions or $N_f=\frac{1}{2}$ while other copies are all projected away, starting with a lattice model with symmetry of $H=G_p \ltimes U_{EM}$.

For our studies of the surface realization of TQCPs, we have elevated the number of fermions in the lattice model from $N_f=\frac{1}{2}$ in $3D$ to $N_f=1$ in $4D$ so that the $4D$ lattice model can have a higher lattice symmetry, i.e. an emergent symmetry of $U(1)$ as its lattice symmetry. The $3D$ surface projection operator is therefore simply $\Sigma_z$ rather than $\gamma$ itself to reduce $N_f$ from its bulk value of $N_f=1$ to its surface value $N_f=\frac{1}{2}$.
To ensure single copy of such a theory, we have also further facilitated our discussions by effectively setting our model to be equivalent to a strong $4D$ topological insulator. 



The $4D$ model with lattice symmetry $H=Z^T_2 \ltimes U(1)$ and $N_f=1$ constructed in this subsection can be shown to be dual to the $4D$ topological insulator model studied in Ref.\cite{Qi08} (See also general discussions in Ref.\cite{Fu07,Moore07}) as both have $U(1)$ symmetry as its lattice symmetry. Dualities between different $3D$ gapless topological states were previously discussed in Ref.\cite{Zhou23}. Below we employ a similar strategy to illustrate a duality between a gapped $4D$ topological insulator and the 4D lattice model above.

\subsection{Dualities between the elevated $4D$ lattice model and $4D$ Topological insulators in AII class}

To establish a duality between the $4D$ lattice model we put forward above for the study of $3D$ TQCP and $4D$ strong topological insulators in AII class, we first define
$\chi^T_m=(\chi_{m1},\chi_{m2},\chi_{m3},\chi_{m4})$, $m=\pm$ {\em uniquely} as

\begin{eqnarray}
&& \chi_{m1} ({\bf r} )=\frac{1}{\sqrt{2}} [ \psi_{m\uparrow} ({\bf r}) + \psi^\dagger_{m\uparrow} ({\bf r})],  \nonumber \\
&& \chi_{m2} ({\bf r} )=\frac{1}{\sqrt{2}} [ \psi_{m\downarrow} ({\bf r}) + \psi^\dagger_{m\downarrow} ({\bf r})], \nonumber \\
&& \chi_{m3} ({\bf r} )=\frac{i}{\sqrt{2}} [ \psi_{m\uparrow} ({\bf r}) - \psi^\dagger_{m\uparrow} ({\bf r})],  \nonumber \\
&& \chi_{m4} ({\bf r} )=\frac{i}{\sqrt{2}} [ \psi_{m\downarrow} ({\bf r}) - \psi^\dagger_{m\downarrow} ({\bf r})].
\label{fixing}
\end{eqnarray}
Here ${1,2,3,4}$ are indices for the real fermions; $\uparrow, \downarrow$ are ones for spins and $m=1,2$ are states with orbitals of $\Sigma_z =\pm 1$. $\psi^\dagger_{m\uparrow,m\downarrow}$, 
($\psi_{m\uparrow,m\downarrow}$) are
the creation (annihilation) operators of complex physical fermions with given orbitals $\pm$ and spins $\uparrow,\downarrow$. The eight component real fermions form a fermionic representation of $Spin(8)$ group.
For a 3D TQCP in DIII class topological states with $N_f=\frac{1}{2}$ fermions, the real fermion representation is the one of $Spin(4)$ group as studied in Ref.\cite{Zhou23}.

The 4D lattice model with $N_f=1$ is elevated out of a $3D$ DIII class topological superconductor.
In the real fermion scheme above, it shall have the following explicit matrix operator structure in Eq.\ref{4D},

\begin{eqnarray}
\alpha_x&=&\tau_z \otimes \sigma_z, \alpha_y=-\tau_z \otimes \sigma_x, \alpha_z=\tau_x\otimes 1\nonumber \\
\alpha_\eta&=&\tau_y \otimes \Sigma_y, \beta_1=\tau_y \otimes \Sigma_x, q^D_1=\tau_x \otimes \sigma_y \otimes \Sigma_z . \nonumber \\
&& [\alpha_i,  q^D_1] =[\beta_1, q^D_1], i=x,y,z,\eta.
\label{abq1}
\end{eqnarray}
Notice again $\alpha_i$, $i=x,y,z,\eta$ are real symmetric matrices and $\beta_1=\beta_0\otimes \Sigma_x $ is anti-symmetric purely imaginary.

A $4D$ topological insulator on the other hand has the following structure in Eq.\ref{4D} in terms of real fermions defined in Eq.\ref{fixing},

\begin{eqnarray}
\alpha_x&=&\Sigma_z \otimes \sigma_x, \alpha_y=\Sigma_z \otimes \sigma_z, \alpha_z=-\tau_y \otimes \Sigma_z \otimes \sigma_y, \nonumber \\
\alpha_\eta&=&\tau_y \otimes \Sigma_y, \beta_1=\tau_y \otimes \Sigma_x, q_1=\tau_y \otimes 1 \otimes 1 \nonumber \\
&&[\alpha_i, q_1] =[\beta_1, q_1], i=x,y,z,\eta.
\label{abq2}
\end{eqnarray}
where $q_1$ here and $q^D_1$ above with the superscript inferring the dual theory are the operators defining the $U(1)$ lattice symmetry in Eq.\ref{abq2} and Eq.\ref{abq1} respectively.

We can establish a duality between the two physical phenomena by noticing the $U(1)$ symmetry operator in the elevated $4D$ lattice model for TQCPs and the one in a $4D$ topological insulator, are part of a $spin(3)$ subgroup given by

\begin{eqnarray}
\tau_x \otimes \Sigma_z \otimes \sigma_y,
\tau_y, \tau_z \otimes \Sigma_z \otimes \sigma_y.
\end{eqnarray}  

Define a Spin(3) rotation that connects two $U(1)$ charges as

\begin{eqnarray}
U_0&=& \exp( i\frac{\pi}{3} \int \chi^T Q_0 \chi), \chi^T=(\chi^T_+,\chi^T_-); \nonumber \\
Q_0 &= &\frac{1}{2\sqrt{3}} [ \tau_x \otimes \Sigma_z \otimes \sigma_y +
\tau_y + \tau_z \otimes \Sigma_z \otimes \sigma_y ].
\label{Dual}
\end{eqnarray}
We find that two $4D$ theories defined in Eq.\ref{4D} with operators given in Eqs.\ref{abq1},\ref{abq2} respectively are dual to each other and are related by the duality transformations in Eq.\ref{Dual}.

To prove the equivalence, we start with the theory for $4D$ topological insulators in AII class specified in Eq.\ref{abq2}. We then take into account the following relations. For any operators

\begin{eqnarray}
A_i=\int \chi^T \alpha_i \chi, B_1=\int \chi^T \beta_1 \chi,
\end{eqnarray}
where $\alpha_i$, $i=x,y,z,\eta$ and $\beta_1$ are defined in Eq.\ref{abq2},  under the transformations in Eq.\ref{Dual} we can define

\begin{eqnarray}
A_i \rightarrow A'_i=U_0 A_i U_0^{-1}, B'_1 \rightarrow B'_1=U_0 B_1 U_0^{-1}.   
\end{eqnarray}
One finds explicitly that

\begin{eqnarray}
A'_i=\int \chi^T  \alpha^D_i \chi, B'_1=\int \chi^T \beta^D_1 \chi, \nonumber \\
\alpha^D_{x,y,z}=q_1^D q_1 \alpha_{x,y,z}, \alpha^D_\eta=\alpha_\eta,  \beta^D_1 =\beta_1.
\label{Duality}
\end{eqnarray}

One can verify explicitly that $\alpha^D_i$, $i=x,y,z,\eta$ and $\beta^D_1$ obtained in Eq.\ref{Duality} are equivalent to the operators defined in the elevated $4D$ lattice model in Eq.\ref{abq1} that we have employed for a boundary representation of a $3D$ TQCP in DIII class topological states. This proves the duality between the two models.

\subsection{Gapless states in $(d+1)$-dimension theories: More emergent symmetries}

Before leaving this section, we briefly comment on $U^{(d+1)}_{EM}=\tilde{U}_{D}(\lambda)$, the emergent symmetry at gapless TQCPs in the $(d+1)$-dimension lattice model with protecting symmetry $H=Z^T_2 \ltimes U(1)$, higher than the 
protection symmetry $G_p=Z^T_2$ in the $d$-dimension lattice model we start with. At gapless TQCPs in the later model ($d$-dimension), the emergent symmetry is $U_{EM}=U(1)$.

$N_f=1$ Dirac fermions have eight components in the real fermion representation and the unitary transformations in the real representation form an $Spin(8)$ group.
$Spin(8)$ has 28 generators all of which have to be antisymmetric and purely imaginary hermitian in the real fermion representation.

One can verify that the emergent symmetry in the $(d+1)$-dimension is simply an $Spin(3)$ subgroup generated by

\begin{eqnarray}
Q_x&=&\int d{\bf r} \chi^T ({\bf r})\gamma \otimes \Sigma_x \chi({\bf r}), \nonumber \\
Q_y&=&  \int d{\bf r} \chi^T ({\bf r})\Sigma_y \chi({\bf r}), \nonumber \\
Q_z&=&\int d{\bf r} \chi^T ({\bf r})\gamma \otimes \Sigma_z \chi({\bf r}), \gamma=i \alpha_x\alpha_y\alpha_z.
\end{eqnarray}

The corresponding $Spin(3)$ group leads to $\tilde{U}_{D} (\lambda)$, the full deformation group in the $(d+1)$-dimension lattice.
The gapless TQCPs in $d+1$-dimension lattices are symmetric under the full deformation  group, in addition to the deformation subgroup $U_{d+1} \subset \tilde{U}_{D}(\lambda)$ discussed before.
The emergent symmetry in this case is given by $U^{(d+1)}_{EM}=\tilde{U}_{D} (\lambda)$. 
Note that the lattice $U(1)$ symmetry group in the $d$-dimension lattice model is generated by $\gamma$, without depending on $\Sigma_{x,y,z}$.

The emergent infrared symmetry, together with the $(d+1)$-dimension protecting symmetry $H$ group, lead to the full symmetry group that leaves $(d+1)$-dimension gapless TQCPs in the lattice model invariant. That is

\begin{eqnarray}
U_{EM}^{(d+1)}=Z^T_2 \ltimes Spin(3) \mbox{ or $Z^T_2 \ltimes SU(2)$}
\end{eqnarray}
much higher than the lattice symmetry group of $H=Z^T_2 \ltimes U(1)$ for gapless TQCPs in $d$-dimensions.

\section{Renormalization group equations for a ring of fixed points}

In this appendix, we outline the solutions to the renormalization group equation (RGE) near a TQCP. We show that by forcing TQCP strongly interacting with protecting symmetry,
we need to identify the TQCP with the protecting symmetry $G_p=Z^T_2$ with a scale invariant time reversal symmetric fixed point among a ring of fixed points. The ring of fixed points, all without protecting symmetries except at $g_{Y2}=0$ pionts , can be obtained by taking into account the anti-commuting relations between $\beta_{1,2}$ and $\alpha_i$, $i=1,2,3$
as well as
\begin{eqnarray}
Tr \beta_1 \beta_2=Tr \beta^3_1\beta_2=Tr \beta_1 \beta^3_2=0, \{ \beta_1,\beta_2 \}=0.
\end{eqnarray}
Diagrammatically,  the RGE can be obtained using the diagrams in Fig.\ref{RGED}.

Eq.\ref{RGE} can be reorganized into
\begin{eqnarray}
\frac{d\tilde{g}_{x,y}}{d t} &=& \tilde{g}_{x,y} ( \frac{d-3}{2} +\tilde{g}^2_{Y}), \nonumber \\
\frac{d\tilde{g}^2_{Y}}{d t} &=&2 \tilde{g}^2_{Y} (\frac{d-3}{2} +\tilde{g}^2_{Y}), \nonumber \\
\frac{d\tilde{g}_4}{d t} &=&(d-3) \tilde{g}_4 +\tilde{g}^2_4 - c\tilde{g}^2_{Y}. 
\label{RGE1}
\end{eqnarray}

Here we have introduced 
\begin{eqnarray}
&& \tilde{g}_{x}= \tilde{g}_{Y1} \cos\theta +\tilde{g}_{Y2} \sin\theta, \nonumber \\
&& \tilde{g}_y=-\tilde{g}_{Y1} \sin\theta +\tilde{g}_{Y2}\cos\theta, \theta\in [0,2\pi] \nonumber \\
&& \tilde{g}^2_Y=\tilde{g}^2_{Y1} +\tilde{g}^2_{Y2}.
\end{eqnarray}
One can see clearly that Eq.\ref{RGE1} is invariant with respect to $SO(2)$ rotations in the plane of $g_{Y1}-g_{Y2}$ and is independent of the choice of $\theta$-angle.

For $d<3$ or dimensions lower than three, the solutions to Eq.\ref{RGE} and the renormalization flows are fully determined by the lines defined as
\begin{eqnarray}
\tilde{g}_{x,y}=0 
\end{eqnarray}
and a ring of fixed points defined as
\begin{eqnarray}
\tilde{g}^2_Y=\tilde{g}^2_{Y1}+\tilde{g}^2_{Y2}=\frac{3-d}{2}.
\end{eqnarray}
That is all the renormalization flows are along the radial directions in the plane of $g_{Y1}-g_{Y2}$ towards the ring and and all are terminated at the ring.

The CFTs of these fixed points all break the protecting symmetry $G_p$ except at two points of $\tilde{g}_{Y2}=0$. 
However, for $d> 3$, the ring shrinks to a point at $\tilde{g}^2_Y=0$ and the fixed point is a fully isolated one with the protecting symmetry $G_p=Z^T_2$.


\begin{thebibliography}{References}

\bibitem{Ji20} W. Ji and X.-G. Wen, Phys. Rev. Research {\bf 2}, 033417 (2020).
\bibitem{Verresen21} R. Verresen, Ryan Thorngren, N. Jones and F. Pollmann, Phys. Rev. X {\bf 11}, 041059 (2021).

\bibitem{Thorngren21} R. Thorngren, A. Vishwanath, and R. Verresen, Phys. Rev. B {\bf 104}, 075132 (2021).
 
\bibitem{Ji22}W. Ji, N. Tantivasadakarn, and X.-G. Wen, ArXIv:{\bf 2212.09754}.

\bibitem{Chatterjee22}A. Chatterjee, W. Ji and X.-G. Wen, ArXIv:{\bf 2212.14432}.



\bibitem{Jiang18} H.-C. Jiang, Z.-X. Li, A. Seidel, and D.-H. Lee, Science Bulletin {\bf 63}, 753 (2018).


\bibitem{Verresen18} R. Verresen, N. G. Jones, and F. Pollmann, Phys. Rev. Lett. {\bf 120}, 057001 (2018).


\bibitem{Scaffidi17}T. Scaffidi, D. E. Parker, and R. Vasseur, Physical Review X {\bf 7}, 041048 (2017).


\bibitem{Wen17} X.-G. Wen, Rev. Mod. Phys. {\bf 89}, 041004 (2017). 

\bibitem{Ruhman15} J. Ruhman, E. Berg, and E. Altman, Phys. Rev. Lett. {\bf 114}, 100401 (2015).




\bibitem{Keselman15} A. Keselman and E. Berg, Phys. Rev. B {\bf 91}, 235309 (2015).


\bibitem{Iemini15}  F. Iemini, L. Mazza, D. Rossini, R. Fazio, and S. Diehl, Phys. Rev. Lett. {\bf 115}, 156402 (2015).



\bibitem{Kainaris15} N. Kainaris and S. T. Carr, Phys. Rev. B {\bf 92}, 035139 (2015).




\bibitem{Baum15} Y. Baum, T. Posske, I. C. Fulga, B. Trauzettel, and A. Stern, Phys. Rev. Lett.{\bf 114}, 100401 (2015).



\bibitem{Fidkowski11a}L. Fidkowski, R. M. Lutchyn, C. Nayak and M. P. A. Fisher, Phys. Rev. B {\bf 84}, 195436 (2011).

\bibitem{Sau11} J. D. Sau, B. I. Halperin, K. Flensberg, and S. Das Sarma, Phys. Rev. B {\bf 84}, 144509 (2011).





 \bibitem{Schnyder08} A. P. Schnyder, S. Ryu, A. Furusaki, and A. W. W. Ludwig, Phys. Rev. B {\bf 78}, 195125 (2008).

\bibitem{Kitaev09} A. Kitaev, AIP Conference Proceedings {\bf1134}, 22 (2009)




 \bibitem{Chen10} X. Chen, Z.-C. Gu, and X.-G. Wen, Phys. Rev. B {\bf 82}, 155138 (2010); Phys. Rev. B {\bf 84}, 235128 (2011).


\bibitem{Chen13} X. Chen, Z.-C. Gu, Z.-X. Liu, and X.-G. Wen, Phys. Rev. B {\bf 87}, 155114 (2013).


 \bibitem{Chen12} X. Chen, Z.-C. Gu, Z.-X. Liu, and X.-G. Wen, Science {\bf 338}, 1604 (2012).
 
 \bibitem{Hastings05} M. B. Hastings, and X.-G. Wen, Phys. Rev. {\bf B} 72, 045141 (2005).
 









\bibitem{Qi11} X.-L. Qi and S.-C. Zhang, Rev. Mod. Phys. {\bf 83}, 1057 (2011).

\bibitem{Hasan10} M. Z. Hasan and C. L. Kane, Rev. Mod. Phys. {\bf 82}, 3045 (2010).

\bibitem{Bernevig13} B. A. Bernevig and T. L. Hughes, {\it Topological insulators and topological superconductors} (Princeton University Press, 2013).

\bibitem{Qi10} X.-L. Qi, T. L. Hughes, and S.-C. Zhang, Phys. Rev. B {\bf 81}, 134508 (2010)
X.-L. Qi, T. L. Hughes, S. Raghu, and S.-C. Zhang, Phys. Rev. Lett. {\bf 102}, 187001 (2009).


\bibitem{Fidkowski10} L. Fidkowski and A. Kitaev, Phys. Rev. B {\bf 81}, 134509 (2010).

\bibitem{Fidkowski11} L. Fidkowski and A. Kitaev, Phys. Rev. B{\bf 83}, 075103 (2011).

\bibitem{Fidkowski13} L. Fidkowski, X. Chen and A. Vishwanath, Phys. Rev. X {\bf 3}, 041016 (2013).


\bibitem{Metlitski15} M. Metlitski, C. L. Kane and M. P. A. Fisher, Phys. Rev. B {\bf 92}, 12511 (2015).

\bibitem{Wang15} C. Wang and T. Senthil, Phys. Rev. X {\bf 5}, 041031 (2015); C. Wang, A. Potter and T. Senthil, Science {\bf 343}, 6171 (2014).

\bibitem{Song17} H. Song, S.-J. Huang, L. Fu and M. Hermele, Phys. Rev. X {\bf 7}, 011020 (2017).


\bibitem{Levin06} M. Levin and X-G. Wen, Phys. Rev. Lett. {\bf 96}, 110405 (2006).


\bibitem{Kitaev06} A. Kitaev and J. Preskill, Phys. Rev. Lett. {\bf 96}, 110404 (2006).











\bibitem{t'Hooft76} G. t'Hooft, Phys. Rev. D {\bf 14}, 3432 (1976).
\bibitem{Adler69} S. L. Adler, Phys. Rev. {\bf 117}, 2426 (1969).
\bibitem{Bell69} J. S. Bell, R. Jackiw, Nuovo. Cimento A {\bf 60}, 47(1969).

\bibitem{Wen13} X.-G. Wen, Phys. Rev. {\bf D} 88, 045013 (2013).





\bibitem{Senthil04} T. Senthil, A. Vishwanath, L. Balents, S. Sachdev and M. Fisher, Science {\bf 303}, 1490 (2004);
T. Senthil, L. Balents, S. Sachdev, A. Vishwanath and M. Fisher, Phys. Rev. B{\bf 70}, 144407 (2004).

\bibitem{Bi19} Z. Bi and T. Senthil, Phys. Rev. X {\bf 9}, 021034 (2019). 
 

\bibitem{Nat21} N. Tantivasadakarn, R. Thorngren, A. Vishwanath, R. Verresen, ArXiv: 2112.01519. 
 


\bibitem{Wen02} X. G. Wen and A. Zee, Phys. Rev. B {\bf 66}, 235110 (2002). 


\bibitem{Beri10} B. B{\'e}ri, Phys. Rev. B {\bf 81}, 134515 (2010).

\bibitem{Kobayashi14} S. Kobayashi, K. Shiozaki, Y. Tanaka, and M. Sato, Phys. Rev. B {\bf 90}, 024516 (2014).


\bibitem{Schnyder11} A. P. Schnyder and S. Ryu, Phys. Rev. B {\bf 84}, 060504(R) (2011).

\bibitem{Matsuura13} S. Matsuura, P. Y. Chang, A. P. Schnyder and S. Ryu, New Jour. Phys. {\bf 15}, 065001 (2013). 












\bibitem{Armitage18} N. P. Armitage, E. J. Mele, and A. Vishwanath, Rev. Mod. Phys. {\bf 90}, 015001 (2018).

\bibitem{Yang21} F. Yang and F. Zhou, Phys. Rev.  {B}  {\bf 103}, 205126 (2021).



\bibitem{Zhou23} F. Zhou, Phys. Rev. B {\bf 107}, 134517 (2023).


\bibitem{Nielson81} H. B. Nielson and M. Ninomiya, Phys. Lett. B {\bf 105}, 219 (1981).
\bibitem{Friedan82} D. Friedan, Commun. Math. Phys. {\bf 85}, 481(1982)
\bibitem{Kaplan92}D. Kaplan, Phys. Lett. B{\bf 288}, 342 (1992). 


\bibitem{Qi08} X.-L. Qi, T. L. Hughes and S.-C. Zhang, Phys. Rev. B{\bf 78}, 195424 (2008).

\bibitem{Fu07} L. Fu, C. L. Kane and E. J. Mele, Phys. Rev. Lett. {\bf 98}, 106803 (2007).

\bibitem{Moore07} J. E. Moore and L. Balents, Phys. Rev. B{\bf 75}, 121306 (2007).


\bibitem{Gross74} D. J. Gross and A. Neveu, Phys. Rev. D {\bf10}, 3235 (1974).

\bibitem{Fei16} L. Fei, S. Giombi, I. R. Klebanov and G. Tarnopolsky, arXiv:1607.05316 and references therein.

\bibitem{Lee07} S. S. Lee, Phys. Rev. B {\bf 76}, 075103(2007).

\bibitem{Balents98} L. Balents, M. P. A. Fisher and C. Nayak, Int. J. Mod. Phys. B {\bf 12}, 1033 (1998).



\bibitem{Zerf16} N. Zerf, C. H. Lin, and J. Maciejko, Phys. Rev. B 94, 205106 (2016).
 \bibitem{Jian17} S.-K. Jian, C.-H. Lin, J. Maciejko, and H. Yao, Phys. Rev. Lett. {\bf 118}, 166802 (2017).
 \bibitem{Zhou22} F. Zhou, Phys. Rev. B {\bf 105}, 014503 (2022).
 
 \bibitem{Grover14} T. Grover, D. N. Sheng and A. Vishwanath, Science {\bf 344}, 280 (2014).















\bibitem{Pollman12} F. Pollman, E. Berg, A. Turner and M. Oshikawa, Phys. Rev. B{\bf 85}, 075125 (2012).

\bibitem{Z_2}When $m_2$ is nonzero, the specific $Z^T_2$ defined above is broken, However, there can be a modified $Z^T_2$ symmetry in this model so that overall symmetry $G_S$ can remain the same.



 
  
  





 \end{thebibliography}
\end{document}